\documentclass[smallextended,numbook]{svjour3}

\journalname{}

\usepackage{ulem}

\usepackage{graphicx,float,amsmath,amssymb,mathrsfs}

\usepackage[english,francais]{babel}

\usepackage{url,hyperref}  

\usepackage[usenames,dvipsnames]{color}

\def\Xint#1{\mathchoice
{\XXint\displaystyle\textstyle{#1}}%
{\XXint\textstyle\scriptstyle{#1}}%
{\XXint\scriptstyle\scriptscriptstyle{#1}}%
{\XXint\scriptscriptstyle\scriptscriptstyle{#1}}%
\!\int}
\def\XXint#1#2#3{{\setbox0=\hbox{$#1{#2#3}{\int}$}
\vcenter{\hbox{$#2#3$}}\kern-.5\wd0}}

\def\dashint{\Xint-}

\renewcommand{\leq}{\leqslant}
\renewcommand{\geq}{\geqslant}

\def\eqdef{:=}
\def\eqlaw{\stackrel{\mbox{\tiny (law)}}{=}}     
\newcommand{\ket}[1]{|\kern.3ex#1\kern.3ex\rangle}
\newcommand{\bra}[1]{\langle\kern.3ex #1 \kern.3ex|}
\newcommand{\mean}[1]{\mathbb{E}\left(#1\right)}
\newcommand{\smean}[1]{\mathbb{E}(#1)}
\newcommand{\cum}[1]{\overline{#1}^c}

\newcommand{\im}{\mathop{\mathrm{Im}}\nolimits}      
\renewcommand{\min}[2]{\mathop{\mathrm{min}}\nolimits\left( #1 , #2\right)}

\newcommand{\proba}{\mathop{\mathbb{P}}\nolimits}

\def\I{{\rm i}}                  
\def\D{{\rm d}}                  

\newcommand{\deriv}[2]{\frac{\mathrm{d}#1}{\mathrm{d}#2}}
\newcommand{\derivp}[2]{\frac{\partial #1}{\partial #2}}
\newcommand{\derivf}[2]{\frac{\delta #1}{\delta #2}}
\newcommand\antiddots{\mathinner{\mkern2mu\raise1pt\hbox{.}\mkern2mu
    \newline \raise4pt\hbox{.}\mkern2mu\raise7pt\hbox{.}\mkern1mu}}

\def\Absor{\kappa}
\def\pot{\Phi}
\def\IntAbsor{K}
\def\iw{v}  

\def\mphi{\overline{\phi}}

\def\e{\mathrm{e}}

\def\normeF{C}

\def\CsteYves{F_0(0)}

\def\ReturnProba{P}

\def\poisson{\Pi}

\graphicspath{{./FiguresSinaiAbs/}}

\begin{document}

\renewcommand{\labelitemi}{$\bullet$}
\renewcommand{\labelitemii}{$\star$}

\selectlanguage{english}

\title{One-dimensional disordered quantum mechanics and\\Sinai diffusion with random absorbers}

\author{Aur\'elien Grabsch \and Christophe Texier \and Yves Tourigny}

\institute{
A. Grabsch
\at 
\'{E}cole normale sup\'{e}rieure de Cachan,
61 avenue du Pr\'{e}sident Wilson,
94235 Cachan cedex, France \\
\email{aurelien.grabsch@ens-cachan.fr}
\and
C. Texier
\at
Univ. Paris Sud; LPTMS, UMR 8626 and LPS, UMR 8502, CNRS; 91405 Orsay cedex,
France\\
\email{christophe.texier@u-psud.fr}
\and
Y. Tourigny
\at 
School of Mathematics, University of Bristol, Bristol BS8 1TW, United
Kingdom\\
\email{y.tourigny@bristol.ac.uk}
}


\date{October 24, 2013}

\maketitle

{\small
\begin{abstract}
We study the one-dimensional Schr\"odinger equation with a disordered potential of the form
$$ 
  V (x) = \phi(x)^2+\phi'(x) + \Absor(x)
$$
where $\phi(x)$ is a Gaussian white noise with mean $\mu g$ and variance $g$, and $\Absor(x)$ is a random superposition of delta functions distributed uniformly on the real line with mean density $\rho$ and mean strength $\iw$.
Our study is motivated by the close connection between this problem and classical diffusion in a random environment (the Sinai problem) in the presence of random absorbers~: 
$\phi(x)$ models the force field acting on the diffusing particle and $\Absor(x)$ models the absorption properties of the medium in which the diffusion takes place.
The focus is on the calculation of the complex Lyapunov exponent $ \Omega(E) = \gamma(E) - \I \pi N(E) $, where $N$ is the integrated density of states per unit length and $\gamma$ the reciprocal of the localisation length.
By using the continuous version of the Dyson-Schmidt method, we find an exact formula, in terms of a Hankel function, in the particular case where the strength of the delta functions is exponentially-distributed with mean $\iw=2g$.
Building on this result, we then solve the general case--- in the low-energy limit--- in terms of an infinite sum of Hankel functions. Our main result, valid without restrictions on the parameters of the model, is that the integrated density of states exhibits the power law behaviour 
$$ 
  N(E) \underset{E\to0+}{\sim} E^\nu
  \hspace{0.5cm}
  \mbox{where }
  \nu=\sqrt{\mu^2+2\rho/g}
  \:.
$$ 
This confirms and extends several results obtained previously by approximate methods.
\end{abstract}
}

\vspace{0.25cm}

\noindent
{\small
}





\section{Introduction}

Localisation in disordered systems has been intensively studied since Anderson's pioneering paper sixty years ago~\cite{LifGrePas88,Luc92}. Yet there are very few systems for which the calculation of basic quantities such as the localisation length and the density of states can be performed analytically. 
This paper presents some new examples where exact calculations are possible.

\subsection{Sinai diffusion with absorption}
\label{sinaiSubsection}

These examples arise naturally from the study of classical diffusion in a random environment with random absorption or killing. 
To be more precise,
consider a classical one-dimensional diffusion $X_t$ on the real line with infinitesimal (backward) generator
\begin{equation}
  \label{generator}
  \mathscr{G} = \deriv{^2}{x^2} + 2\phi(x) \deriv{}{x} - \Absor(x) \,.
\end{equation}
In this expression, $\phi(x)$ is the drift, and the killing rate $\Absor(x)>0$ describes the absorption properties of the diffusive medium.
We shall be interested in the case where $\phi$ and $\Absor$ are noises arising from two
independent stochastic processes~:
\begin{equation}
  \pot(x) = \int_0^x \phi (u)\,\D u 
  \hspace{0.25cm} \mbox{and} \hspace{0.25cm} 
  \IntAbsor(x) = \int_0^x \Absor(u)\,\D u\,.
  \label{stochasticProcesses}
\end{equation}
These processes describe the {\it environment} in which the diffusion takes place.
The particular case where $\pot$ is a Brownian motion--- i.e. $\phi$ is a Gaussian white noise with mean $\mu g$ and variance $g$--- and $\IntAbsor$ is identicallly
zero corresponds to
the ``continuum'' version of the random walk in a random environment introduced by Sinai \cite{Sin82}~; interest in the Sinai model spans many fields of science, ranging from 
mathematics and finance \cite{GemYor93,Shi01}, to statistical physics  \cite{BouComGeoLeD90,LeDMonFis99,MonLeD02}, polymer physics \cite{LubNel02,OshRed09} and population dynamics \cite{AraTsiVin97}. 

The effects of absorption in the presence of a random force field were examined recently  in \cite{HagTex08,LeD09,TexHag09}.
The situation of physical interest is that where the process $\IntAbsor$ is {\it non-decreasing}, i.e. $\Absor$ is positive. It is also natural to ask that  $\IntAbsor$, like the Brownian process $\pot$, be {\it Markovian} and have {\it stationary increments}. This second condition ensures that $\Absor$, like $\phi$, is uncorrelated in space and that its distribution is translation invariant.
These natural requirements are fulfilled automatically when $\IntAbsor$ is a {\it L\'{e}vy subordinator}. We shall, in \S \ref{sec:Models}, go over the aspects of  L\'{e}vy processes that are most relevant to our study.
For the moment, suffice it to say that the absorption rate will take the form
\begin{equation}
  \label{absorptionProcess}
  \Absor(x) = \sum_{n} v_n \, \delta ( x- x_n )
\end{equation}
where the positions $\{x_n\}$ of the \og absorbers \fg{} (or \og impurities \fg{}) are uncorrelated and distributed uniformly with a mean density $\rho$. 
The weight $v_n>0$ measures the effectiveness of the absorption at the impurity located at $x_n$. 
The probability density of the weights will be denoted by $p$ and will be  supported on the positive half-line; the mean value  will be denoted by $v >0$.
The resulting absorption process $\IntAbsor$ is a particular kind of L\'{e}vy subordinator known as ``compound Poisson process'' \cite{App04}. 

As is well-known, 
the infinitesimal generator gives analytical access to the most important properties of the diffusion $X_t$.  To illustrate this point,
define the transition kernel $P_{\pot,\IntAbsor}(x,y;t)$, conditional on the environment, via~\footnote{
  $\proba(A)$ denotes the probability of event $A$ occuring.
}
\begin{equation}
  \label{transitionKernel}
  {\mathbb P}_x \left ( X_t \in dy\, | \,\pot,\IntAbsor \right ) = P_{\pot,\IntAbsor}(x,y;t)\,\D y\,.
\end{equation}
Here, $x$ denotes the starting point of the diffusion, i.e. $X_0=x$, and $dy$ denotes an interval of infinitesimal length $\D y$ centered
on $y$. For instance, given the environment, the density of the probability that the particle returns at time $t$ to its
starting point, and the probability that the particle survives beyond time $t$ (i.e. is not absorbed before time $t$) are given respectively by
$$
P_{\pot,\IntAbsor} (x,x;t) \;\; \mbox{and} \;\; \int_{\mathbb R} P_{\pot,\IntAbsor} (x,y;t)\,\D y\,.
$$
Furthermore $P_{\pot,\IntAbsor}(\cdot,y;\cdot)$ solves the backward Fokker--Planck (forward Kolmogorov) equation
\begin{equation}
  \frac{\partial u}{\partial t} = {\mathscr G} u\,, \hspace{0.25cm} x \in {\mathbb R}\,,\;\;t >0\,,
  \label{backwardFokkerPlanckEquation}
\end{equation}
subject to the initial condition $u(x,y;0) = \delta(x-y)$.  Since the environment is random, these quantities are themselves random variables, and
the difficulty is to compute expectations over the environment.

\subsection{One-dimensional Schr\"odinger equation with disorder}
\label{disorderedSystem}

It turns out that the (negative of the) generator ${\mathscr G}$ is conjugate to the quantum Hamiltonian
\begin{equation}
  \label{eq:hamiltonian}
  \boxed{ {\mathscr H} = -\deriv{^2}{x^2} + \phi(x)^2 + \phi'(x) + \Absor(x) }
\end{equation}
Indeed, the generator may be writen as 
\begin{equation}
  {\mathscr G} = \e^{-2\pot(x)}\deriv{}{x}\e^{2\pot(x)}\deriv{}{x} + \Absor(x)
\end{equation}
and so it is readily verified that
\begin{equation}
  \label{conjugacy}
  \e^{\pot(x)} \left ( -{\mathscr G} \right ) \e^{-\pot(x)} = {\mathscr H}\,.
\end{equation}
These operators
are therefore spectrally equivalent and we may study the diffusion through the 
quantum system associated with ${\mathscr H}$. For a deterministic environment, this approach is well-illustrated by the work
of Truman {\it et al.} \cite{TruWilYu95}~; for a random environment--- and hence a disordered quantum system--- the idea
was exploited by Bouchaud {\it et al.} \cite{BouComGeoLeD90}.

Let us comment briefly on the structure of the Hamiltonian~: in the absence of the term $\Absor(x)$, it possesses  some underlying symmetry --- the so-called {\it supersymmetry} \cite{Jun96,ComTex98} 
--- encoded in the factorisation 
\begin{equation}
  \label{eq:Hsusy}
  {\mathscr H}_\mathrm{susy} 
 = {\mathscr Q}^\dagger {\mathscr Q}
\end{equation}
where
\begin{equation}
 {\mathscr Q} := -\e^{\pot(x)}\deriv{}{x} \e^{-\pot(x)}  =-\deriv{}{x}+\phi(x)
 \:.
\end{equation}
The function $\e^{\pot(x)}$ appearing in the transformation (\ref{conjugacy}) then has a clear physical meaning~: in the quantum mechanical formulation, it is a zero mode of the supersymmetric Hamiltonian, 
$
{\mathscr H}_\mathrm{susy} \,\e^{\pot(x)}=0\,.
$
In the classical diffusion problem, its square $\e^{2\pot(x)}$, if normalisable, is the equilibrium distribution 
\cite{Gar89}. 
In the quantum Hamiltonian (\ref{eq:hamiltonian}), the term $\Absor(x)$ breaks the supersymmetry.

Since our model is a mixture of two noises, it will be helpful to use the following
terminology to
distinguish between three different kinds of potentials:
\begin{enumerate}
\item $\phi^2+\phi'$ ({\it supersymmetric})~; here, $\phi$ will be a Gaussian white noise.
\item $\kappa$ ({\it scalar})~; here, $\kappa$ will be a positive (non-Gaussian) white noise.
\item $\phi^2+\phi' + \kappa$ ({\it mixed}), with $\phi$ and $\kappa$ as above.
\end{enumerate}
Potentials that are either scalar or supersymmetric will be called {\it monolithic}.
Further details on the nature of the noises $\phi$ and $\kappa$ will be given in \S \ref{sec:Models}.

Monolithic quantum Hamiltonians involving L\'evy noises have received some attention in the literature. In the particular case where the L\'{e}vy
process is a subordinator--- possibly with a singular L\'{e}vy measure--- Kotani~\cite{Kot76} and Comtet {\it et al.}~\cite{ComTexTou11} have studied 
the low-energy spectral density 
of scalar and supersymmetric Hamiltonians  respectively.
Bienaim\'{e} \& Texier~\cite{BieTex08} considered monolithic disorder arising from compound Poisson processes with jump heights that have an infinite
variance~; they observed some unusual localisation properties such as superlocalisation
(cf. \cite{BooLuc07} and references therein).

\begin{figure}[!ht]
  \centering
 \includegraphics[width=0.475\textwidth]{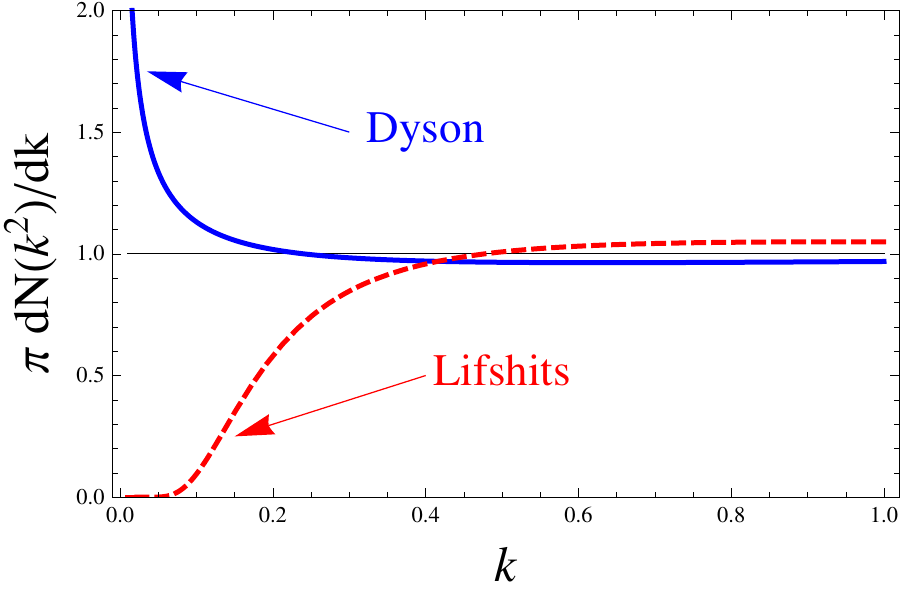}
 \hfill
 \includegraphics[width=0.475\textwidth]{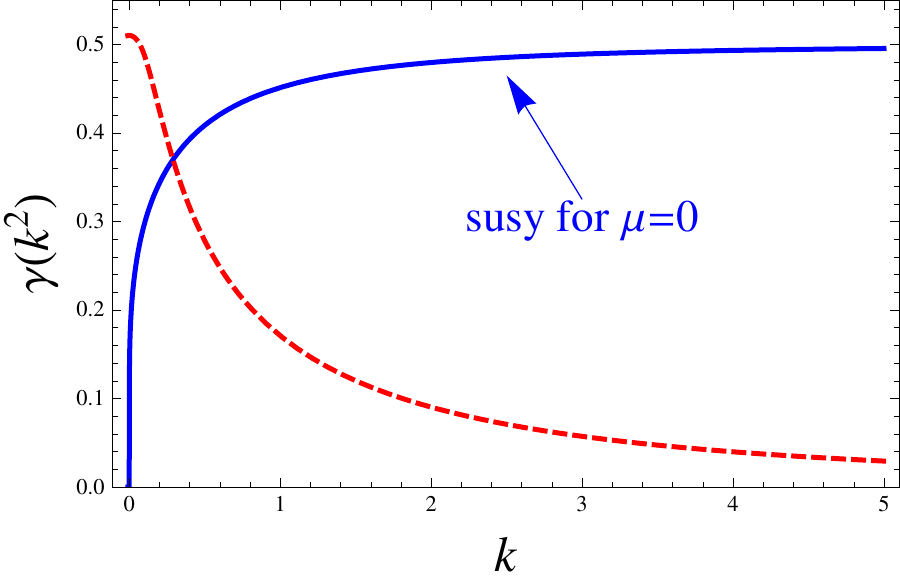}
  \caption{
   {\it 
   The spectral density and the Lyapunov exponent as functions of the wave number:
   comparison between monolithic ``supersymmetric'' disorder $\phi$ with $g=1$ and $\mu=0$ (solid blue lines), and 
   monolithic ``scalar'' disorder $\kappa$ with $\rho=0.2$ and $v=5$ (red dashed lines).
   }
   }
  \label{fig:ComparisonSusyNieuwenhuizen}
\end{figure}

Our motivation for studying the mixed case, where {\it both} supersymmetric and scalar disorders are present,
is to understand the interplay between two types of disorder--- each of them, when taken separately, resulting in quite different phenomenologies. This is illustrated in Figure~\ref{fig:ComparisonSusyNieuwenhuizen}. This theme has already been touched upon in a paper by Hagendorf \& Texier \cite{HagTex08}, who studied the case where $\Absor$ and $\phi$ are Gaussian white noises.
The case of correlated Gaussian white noises was also discussed more recently in connection with products of random matrices in $\mathrm{SL}(2,\mathbb{R})$~\cite{ComLucTexTou13}.

One concrete goal of the present study is to settle the conjecture that, for the mixed model,
\begin{equation}
  N(E) \underset{E\to0+}{\sim} E^\nu
  \hspace{0.5cm}
  \mbox{where }
  \nu = \sqrt{\mu^2 + 2\rho/g}
  \:.
\label{lowEnergyBehaviour}
\end{equation}
For the particular case $\mu=0$, Texier \& Hagendorf~\cite{TexHag09} 
used the ordered spectral statistics~\cite{Tex00,TexHag10} of the Hamiltonian $\mathscr{H}_\mathrm{susy}$ to establish 
the validity of this formula in the limit of a low density of efficient absorbers, i.e. $\rho\to0$ and $v\to\infty$ (cf. Figure~\ref{fig:PhaseDiagram}).
Monte Carlo simulations~\cite{TexHag09} have suggested that this asymptotic formula should remain valid for arbitrary $\rho$. 
By using a real space renormalisation group approach, Le~Doussal \cite{LeD09} found, among many other things, the power law \eqref{lowEnergyBehaviour} in the case $\mu\neq0$,
albeit again with the restriction $\rho \rightarrow 0$ and $v \to \infty$.
The study of a solvable version of the model, free of restrictions on the parameters, 
is therefore of great interest.

\subsection{Outline of the paper}
\label{outlineSubsection}

The focus of the paper is the calculation of the {\it complex Lyapunov exponent}
\begin{equation}
  \label{eq:characteristicFunction}
  \Omega := \gamma - \I \pi N
\end{equation}
associated with the disordered Hamiltonian (\ref{eq:hamiltonian}).
Here, $\gamma$  denotes the usual (real) Lyapunov exponent--- that is, the reciprocal of the localisation length--- and $N$ is the integrated density of states (IDOS) per unit length. Both quantities are self-averaging when the sample size goes to infinity.
Particular emphasis will be put on the study of $N$ because the density of states of the disordered system gives immediate access to the density of the probability that the diffusion returns to its starting point (see \S5 of \cite{BouComGeoLeD90} or \cite{HagTex08})
\begin{equation}
  \label{returnProbability}
  \ReturnProba(t) := \mean{  P_{\pot,K} (x,x;t) } = \int_0^\infty N'(E) \,\e^{-E t}\,\D E
\end{equation}
where $\mean{\cdot}$ denotes averaging with respect to the random processes $\pot$ and $\IntAbsor$.
A secondary aim of the paper is to give a unified description of spectral and localization properties when $\Absor$ and $\phi$ are considered separately or combined.
To this end, we use the Dyson-Schmidt method developed by Frisch \& Lloyd and Halperin in the case $\phi=0$, and, in \S \ref{sec:Methodology},
show how to adapt it to our situation.
In \S \ref{sec:LimitingBehaviours}, we illustrate the approach by recovering straightforwardly the known relevant analytical results for the 
monolithic Hamiltonian (\ref{eq:hamiltonian}). We also study in some detail certain limits which will serve as points of comparison when
we examine the mixed case later on.

The remainder of the paper is devoted to the study of the mixed case,  where
$\pot$ is a Brownian motion with drift $\mu g$ and variance~$g$, whilst
$\IntAbsor$ is a compound Poisson process of intensity $\rho$, with a probability density $p$ and a mean $v$ for the weights.

We begin, in Section~\ref{sec:MixedCaseFixedWeights}, with the case where $\IntAbsor$ is a Poisson process, i.e.
$p(v_n)= \delta(v_n-v)$~; this is the situation originally considered in Refs.~\cite{TexHag09,LeD09}. 
Despite its apparent simplicity, this problem turns out to be a difficult one, and it is only for the particular
case where the energy vanishes that we succeed in making some analytical progress.

Then, in \S \ref{sec:SolvableMixedCase}, we examine the case 
where $\IntAbsor$ is a compound Poisson process with exponentially-distributed jumps. 
One of our main result is the discovery of a new solvable case that arises when the mean value $\iw$ of the jumps is related to the strength $g$ of the force field via $\iw=2g$.

In \S \ref{sec:HankelSeries}, we extend the validity of the asymptotic formula
(\ref{lowEnergyBehaviour}). We begin with the study of the exponentially-distributed case for an arbitrary mean $v$.
We show how,  in the low energy limit, the problem can be solved in terms of a Hankel series whose first few terms
we compute explicitly by recurrence. This Hankel series may in fact be viewed as a perturbation expansion around the
solvable point $v=2g$. We then explain how the same approach works also
in the general case of an arbitrary distribution of the positive jump heights, with no restriction on the mean $v$.
This demonstrates the advantages of the Dyson-Schmidt method over the other approaches of Refs.~\cite{TexHag09,LeD09} which
are inherently limited to ``small'' regions of the parameter space, as illustrated in Figure~\ref{fig:PhaseDiagram}. 

Finally, we elaborate in \S \ref{sec:Conclusion} the implications of our findings for Sinai diffusion with random absorbers.


\section{Methodology}
\label{sec:Methodology}

In this section, we review the aspects of the Dyson-Schmidt method relevant to our particular problem. 
The upshot is that, for the particular kind of disorder that we consider, the calculation of the complex Lyapunov exponent reduces to the solution of a certain linear second order differential equation~; this is Equation~(\ref{eq:FdeQ}) below.

\subsection{Nature of the disorder -- L\'evy noises}
\label{sec:Models}

We call {\it L\'{e}vy process} any Markov process, started at zero, with right-continuous, left-limited paths
and stationary independent increments  \cite{App04}.

The characteristic function of any L\'{e}vy process, say $L(x)$, is an exponential of $x$, i.e.
\begin{equation}
  \label{eq:DefLevyExponent}
  \mean{ \e^{-\I q L(x)} } = \e^{-x\,\Lambda(q)}
  \:.
\end{equation}
Furthermore, the {\it L\'{e}vy exponent} $\Lambda(q)$ can always be expressed in the form
\begin{equation}
  \label{eq:LevyKhintchine}
  \Lambda(q) = \I a\, q + \frac12b\,q^2
  +\int_{\mathbb{R}} m(d y)\,
  \left(
   1 - \e^{-\I q y} - \frac{\I q y}{1+y^2}
  \right)
\end{equation}
for some numbers $a$ and $b$, and some measure $m(d y)$ such that
\begin{equation}
\int_{\mathbb R} m(d y)\,\min{1}{y^2} <\infty\,.
\label{levyMeasure}
\end{equation}
$m(d y)$ is called the {\it L\'{e}vy measure} of the process.
When the measure has support $\mathbb{R}_+$ (as happens for example when $b=0$ and the process is a {\it subordinator}), it must satisfy
\begin{equation}
\int_{\mathbb{R}_+}m(d y)\,\min{1}{y}<\infty
\:.
\label{levyCondition}
\end{equation}

The L\'{e}vy processes that are most relevant to the present study are:
\begin{enumerate}

\item 
  The {\it Brownian motion with drift $a$ and variance $b$}. This corresponds to the case where
  $$
  a\neq0,\; b\neq0 \;\text{and}\; m(d y) = 0\,.
  $$

\item 
  The {\it Poisson process $\poisson(x)$ of intensity $\rho$}. In this case, $a=1/2$, $b=0$ and    $m(d y)$ is the discrete measure with weight $\rho$ at $y=1$~; 
  equivalently
  $$
  \Lambda (q) = \rho\, \left ( 1-\e^{-\I q} \right )
  \,.
  $$
  The corresponding noise takes the form
  $$
  \poisson'(x)  = \sum_{n=1}^\infty \delta(x-x_n)
  $$
where the partition
$$
0 < x_1 < x_2 < \cdots
$$
is such that the spacings $\ell_n=x_{n+1}-x_n$ are independent and exponentially distributed. In other words, for $b > a \geq 0$,
$$
{\mathbb P} \left (a <  \ell_n < b \right ) = \int_a^b \rho \,\e^{-\rho x}\,\D x
\:.
$$
This is the situation that would arise if impurities were dropped independently on the positive half-line uniformly at random with mean density~$\rho$.

\item  
  The  {\it compound Poisson process} 
  $$
 L(x) =  \sum_{n=1}^{\poisson(x)} y_n \,.
  $$
In this expression, the $y_n$ form a sequence of independent identically distributed random variables,
and $\poisson(x)$ is a Poisson process of intensity $\rho$, independent of the $y_n$.
The corresponding L\'evy noise now takes the form of a superposition of  delta-functions with random weights
  \begin{equation}
  \label{eq:LevyNoiseCPP}
  L'(x)  = \sum_{n=1}^\infty y_n\,\delta(x-x_n)
  \:.
  \end{equation}
If we suppose that the distribution of the jumps $y_n$ has a density, say $p$, then the L\'{e}vy exponent is given by
$$
   \Lambda(q)  =
  \rho \int_{\mathbb{R}} \D y\,p(y)\,
  \left( 1 -\e^{-\I q y} \right)
$$
i.e.
\begin{equation}
  \boxed{ \Lambda(q) = \rho \, \left( 1 - \hat p(q) \right)}
  \hspace{0.5cm}\mbox{where}\hspace{0.5cm}
  \hat p(q)=\int\D y\,p(y)\,\e^{-\I qy}
  \:.
\label{eq:LevyExponentRegular}
\end{equation}
\end{enumerate}

Other more sophisticated examples of L\'{e}vy processes may be found in \cite{App04b}, but  they will play no part in what follows.
Additional useful statistical properties of L\'evy noises
are discussed in Appendix~\ref{appendix:SPLN}.

\subsection{The Riccati variable}
\label{sec:Ricatti}

Some of the spectral and localisation properties of the one-dimensional random Hamiltonian (\ref{eq:hamiltonian}) may be obtained by using  the Dyson-Schmidt method~\cite{LifGrePas88,Luc92}.
The starting point is the initial-value problem
\begin{equation}
  \label{cauchyProblem}
  {\mathscr H} \psi(x;E) = E\, \psi(x;E) 
  \hspace{0.25cm} \mbox{subject to $\psi(0;E) = 0$ and $\psi'(0;E)=1$}
  \,.
\end{equation}
The study of the statistical properties of $\psi(x;E)$ is facilitated by the introduction of the Riccati variable 
\begin{equation}
  \label{eq:DefRiccati}
  z(x;E) := \frac{\psi'(x;E)}{\psi(x;E)} - \phi(x)
  \:.
\end{equation}
The Schr\"odinger equation for the wave function $\psi$, translates into the following first-order non linear equation for the Riccati variable:
\begin{equation}
  \label{eq:RicEq}
  \deriv{}{x}z(x,E)=-E-z(x,E)^2 -2z(x,E)\,\phi(x) + \Absor(x)
  \:.
\end{equation}
The scalar random potential $\Absor$  appears as an {\it additive} noise whereas the random superpotential $\phi$ appears as a {\it multiplicative} noise. 
The stochastic differential equation is here to be interpretated in the sense of Stratonovich~\cite{Gar89,Oks10}. 
Roughly speaking, this interpretation amounts to viewing the Brownian process as the limit of some differentiable process~; this is appropriate  in our particular, physically-motivated context.

Throughout the paper, $\Phi$ will be the L\'{e}vy process with exponent
$$
\I \,\mu g \,q + \frac{g}{2} q^2\,.
$$
In other words, $\phi := \Phi'$ will always be a Gaussian white noise of mean $\mu g$ and covariance $g \,\delta (x-x')$.
On the other hand, the process $K$ corresponding to $\Absor(x)$ 
will be the compound Poisson process with L\'{e}vy exponent (\ref{eq:LevyExponentRegular}) 
with $p$ supported on $\mathbb{R}_+$.
The distribution $f(z;x)$ of the Riccati variable then obeys the integro-differential equation~\cite{FriLlo60,BieTex08,ComTexTou10}
\begin{eqnarray}
  \label{FPEForDistributionF}
  \derivp{}{x}f(z;x)
  =&&\derivp{}{z}\left[(E+2\mu g z+z^2)f(z;x)\right]
  + 2g\derivp{}{z}\left[z\derivp{}{z}\left[z f(z;x)\right]\right]
  \nonumber
  \\
  &&+ \rho\int_0^\infty \D y\,p(y)\, \left[ f(z-y;x) - f(z;x) \right]\,.
\end{eqnarray}

The significance of the various terms in this equation is more easily grasped if one thinks of the Riccati equation (\ref{eq:RicEq})
as describing how the ``position'' $z$ of some fictitious particle changes with ``time'' $x$
\cite{FriLlo60,BieTex08,ComTexTou10}.
The first term on the right-hand side of Equation (\ref{FPEForDistributionF}) is then a drift term due to the \og deterministic force \fg{}
acting on the particle.  
The following term is a diffusion term associated with the Brownian part of the supersymmetric noise.
Finally, the integral term comes about because the scalar noise
causes the particle to jump from $z(x_n^-)$ to $z(x_n^-)+v_n$ as it traverses 
an impurity at $x_n$.
The drift part 
$$
\deriv{}{x}z=-E-2 \mu g z-z^2\,,
$$
on its own, induces a flow of the process through $\mathbb{R}$, unless  the force $-E-2 \mu g z-z^2$ vanishes for some real $z$. 
This implies (see ~\cite{LifGrePas88}) that, as $x \to\infty$, the distribution reaches a limit distribution $f(z)$ for a steady current $-N(E)$~:
\begin{align}
  \label{eq:IntegralEquationForF}
  N(E) = & (E+2 \mu g z+z^2)f(z) 
  + 2 g\,z\deriv{}{z}\left[z f(z)\right]
  \nonumber \\
  & - \rho \int_0^\infty \D y\,p(y)\, \int_{z-y}^z\D t \,f(t)\,.
\end{align}
The current can be interpreted as the number of infinitudes per unit length of the process $z(x)$---that is, the averaged density of zeros of $\psi(x;E)$. It has therefore the meaning of the integrated density of states per unit length of the quantum Hamiltonian \eqref{eq:hamiltonian}.

An important concept of the localisation theory, which we
reviewed recently in \cite{ComTexTou13},
is that of \textit{Lyapunov exponent}
\begin{equation}
  \label{eq:DefLyapunov}
  \gamma(E) \eqdef \lim_{x\to\infty} \frac{\ln |\psi(x;E)|}{x}
  \:.
\end{equation}
This self-averaging quantity characterizes the {\it average} growth rate of the envelope of the wave function~; its reciprocal provides a measure of the localisation of the eigenstates.
Assuming some ergodic property for the Riccati variable, one can show that the Lyapunov exponent may be obtained from the stationary distribution~\cite{LifGrePas88,ComTexTou10} via
\begin{equation}
  \label{eq:DefLyapunov2}
  \gamma(E) -\mphi
  = \lim_{x\to\infty}\frac1x\int_0^x\D u\,z(u) 
  = \dashint_{-\infty}^{+\infty}\D z\: z\,f(z)
  \:,
\end{equation}
where we have introduced the notation $\mphi=\mean{\phi(x)}$.
The last integral on the right of this expression is a Cauchy principal value integral.
The fact that, when $N(E)>0$, the integral in the usual sense does not exist is obvious from the Rice formula
$z^2f(z)\simeq N(E)$ as $z \rightarrow \infty$.

\subsection{The Fourier transform}
\label{sec:FdeQ}
Solving the integro-differential equation \eqref{eq:IntegralEquationForF} directly is a difficult task.
An important technical simplication arises by considering instead the equation for
the {\it Fourier transform}
\begin{equation}
\hat{f} (q) := \int_{-\infty}^\infty \D z f(z) \,\e^{-\I q z}\,.
\label{fourierTransform}
\end{equation}
Because we have deliberately restricted our attention to the case where $\Phi$ has no jumps, it turns out that the
equation for $\hat{f}$
takes on a purely {\it differential} form~: 
\begin{equation}
  \label{eq:oldFdeQ} 
  \hspace{-0.3cm} 
  \left[
    - \deriv{}{q}\left(1+2\I\, g\,q\right)\deriv{}{q}
    + 2\I\, \mu g \deriv{}{q}
    + E + \I\frac{\Lambda(q)}{q}
  \right]\hat{f}(q) = 2\pi\,N(E)\,\delta(q)\,.
\end{equation}
Furthermore,
the fact that $f$ is a probability density
has the following elementary implications for  its transform $\hat{f}$:
\begin{enumerate}
\item $\lim_{q\to\infty}\hat{f}(q)=0$~;
\item  $\hat{f}(-q)=\hat{f}(q)^*$~; 
\item  $\hat{f}(0)=1$.
\end{enumerate}
The Dirac delta on the right-hand side of Equation (\ref{eq:oldFdeQ}) translates into a jump condition on the derivative, namely \cite{FriLlo60,Hal65,Kot76}
$$
-\hat{f}'(0^+)+\hat{f}'(0^-)=2\pi\,N(E)\,.
$$
On the other hand, since we have
$$
\gamma(E)-\mphi=-\im[\hat{f}'(0^+)]\,,
$$
we may write 
\begin{equation}
  \label{eq:RelationFPrimeOmega}
  \boxed{
   \Omega(E+\I0^+) = \mphi + \I \,\hat{f}'(0^+)
  }
\end{equation}
where $\Omega$ is the characteristic function defined by Eq.~\eqref{eq:characteristicFunction}.

We have thus reduced the problem of finding $\Omega$ to that of
finding a particular solution of the
{\it one-sided, homogeneous version} of Equation (\ref{eq:oldFdeQ})~: for $q>0$,
\begin{equation}
  \label{eq:FdeQ} 
  \hspace{-0.3cm} 
  \boxed{
  \left[
    - \left(1+2\I\, g\,q\right)\deriv{^2}{q^2}
    + 2\I\, (\mu-1) g \deriv{}{q}
    + E + \I\frac{\Lambda(q)}{q}
  \right]\hat{f}(q) = 0
  }
\end{equation} 
The particular solution we require is the solution that decays to zero as $q \rightarrow +\infty$ and that satisfies $\hat{f}(0)=1$.
Equation (\ref{eq:FdeQ}) is at the heart of our approach, and the remainder of the paper will be occupied with its solution.
The solvable cases will correspond to situations where (\ref{eq:FdeQ}) can be transformed  into the hypergeometric equation.

\subsubsection*{Example 1~: Supersymmetric noise}

As a first illustration, we show how to recover efficiently the result first obtained in Ref.~\cite{BouComGeoLeD90}.
For $\rho=0$, i.e. $\Absor=0$, Equation (\ref{eq:FdeQ}) reduces to  
\begin{equation}
  \label{eq:Eqfsusy}
  \left[
    -(1+2\I g\,q)\deriv{^2}{q^2} + 2\I g (\mu-1)\deriv{}{q} + E
  \right] \hat{f}(q) = 0  
  \:.
\end{equation}
For $E=+k^2$, the solution vanishing at infinity is proportional to a Hankel function:
\begin{equation}
  \hat{f}(q) = c\, (1+2\I g\,q)^{\mu/2} \, 
  H^{(1)}_\mu\left(\frac{k}{g}\sqrt{1+2\I g\,q}\right)
\end{equation}
where $c$ must be chosen so that $\hat{f}(0)=1$.
The characteristic function (\ref{eq:RelationFPrimeOmega}) is therefore a ratio of Hankel functions:
\begin{equation}
  \label{eq:LyapsusyPositiveEnergy}
    \Omega(k^2+\I0^+) = -\mu g 
    + k\, \frac{ H^{(1)}_{\mu+1}\left({k}/{g}\right) }{ H^{(1)}_\mu\left({k}/{g}\right) }
  \:.
\end{equation}
In Ref.~\cite{BouComGeoLeD90}, this formula was derived in two steps:
first by solving the equation (\ref{FPEForDistributionF}) for the probability distribution $f(z)$~; second
by working out the integral on the right-hand side of Equation (\ref{eq:DefLyapunov2}).

\begin{figure}[!ht]
  \centering
 \includegraphics[width=0.475\textwidth]{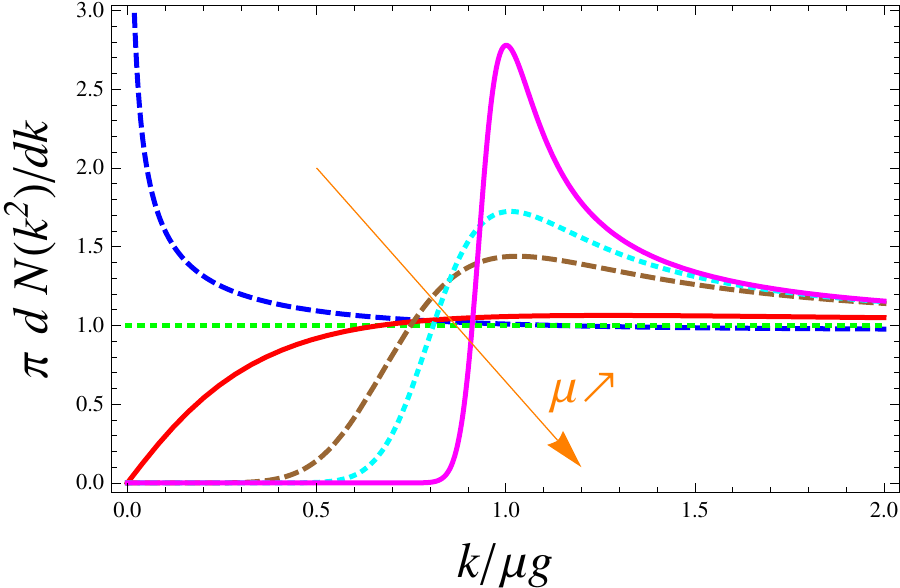}
  \hspace{0.25cm}
  \includegraphics[width=0.475\textwidth]{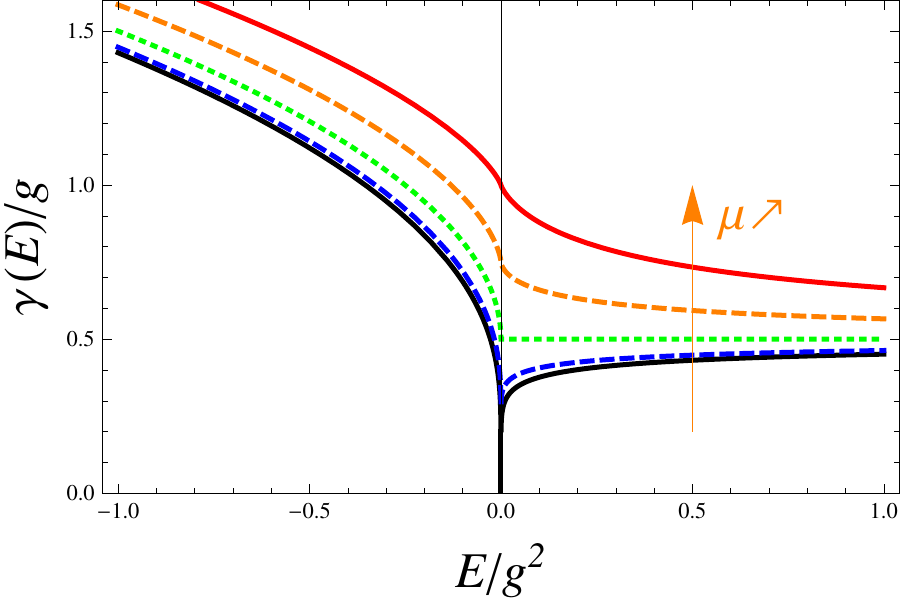}
  \caption{Left~:
   {\it The ratio of the density of states for the supersymmetric model, $N'(k^2)$, to that for the free case, $1/(2\pi k)$, 
   plotted as a function of $k/(\mu g)$ for $\mu=0.25$, $0.5$, $1$, $5$, $10$ and $50$. }
   Right~: {\it Lyapunov exponent for $\mu=0$,
     $0.25$, $0.5$, $0.75$ and $1$.}}
  \label{fig:dos_susy}
\end{figure}

The IDOS and the Lyapunov exponent are plotted in Figure~\ref{fig:dos_susy} for several values of $\mu$. See also Figure~\ref{fig:ComparisonSusyNieuwenhuizen} for the case $\mu=0$.

\subsubsection*{Example 2~: Scalar noise with exponentially-distributed weights}
Let 
\begin{equation}
p(y) = \frac{1}{v}\,\e^{-y/v} \quad \text{for $y \geq 0$}\,.
\label{exponentialDistrution}
\end{equation}
The complex Lyapunov exponent for this case was calculated exactly
by Niewenhuizen in \cite{Nie83}, using an approach different from ours.
For this distribution of the weights, Eq.~(\ref{eq:FdeQ}) becomes
\begin{equation}
  \left[
    - \deriv{^2}{q^2} + E - \frac{\rho}{1/\iw + \I q}
  \right]\hat{f}(q) = 0
  \hspace{0.5cm}\mbox{ for }
  q>0
  \:.
\end{equation}
For $E=-k^2$, the solution vanishing at infinity is the Whittaker function  
\begin{equation}
  \hat{f}(q) = c\, W_{-\frac{\rho}{2k},\frac12}(2k[\I q+1/\iw])
\end{equation}
where $c$ is a normalisation constant. Hence
\begin{equation}
  \label{eq:ResYves}
  \Omega(-k^2)
  = -2 k \,
  \frac{W'_{-\frac{\rho}{2k},\frac12}(2k/\iw)}
  {W_{-\frac{\rho}{2k},\frac12}(2k/\iw)}
\end{equation}
The complex Lyapunov exponent for a positive energy $E=k^2$ is then 
obtained by analytic continuation~; it suffices to replace $k$ by $-\I k$ in the above expression.


\section{Monolithic disorder -- Limiting behaviours}
\label{sec:LimitingBehaviours}


Before we launch into the more difficult study of the mixed models, it is useful to discuss the main physical properties of the two monolithic models of disorder. This section provides a detailed discussion of the limiting behaviours of the integrated density of states and of the Lyapunov exponent.
We shall see in due course that some of the results in the mixed case can be deduced immediately from corresponding results in the monolithic case
after a simple redefinition of the parameters.

\subsection{Supersymmetric disorder}
\label{supersymmetricDisorderSubsection}

The high-energy limit follows from (\ref{eq:LyapsusyPositiveEnergy})~; we find
$$
\Omega(k^2+\I0^+)\underset{k\to\infty}{=}\frac{g}2-\I k+\mathcal{O}(1/k)
\:.
$$

The low-energy behaviour is most conveniently obtained by considering the analytic continuation of 
 (\ref{eq:LyapsusyPositiveEnergy}) to negative values of the energy:
\begin{equation}
  \label{eq:LyapsusyNegativeEnergy}
    \Omega(-k^2) = 
    -k \frac{K'_\mu(k/g)}{K_\mu(k/g)}
\end{equation}
where $K_\mu(z)$ is the MacDonald function~\cite{gragra}. We readily deduce
\begin{align*}
\Omega(E) &\underset{E\to0-}{=} 
  \left\{
  \begin{array}{ll}
   \frac{g}{\ln \left ( \frac{2g}{\sqrt{-E}} \right )-\mathbf{C}} +  \mathcal{O}( E ) 
    & \mbox{for } \mu = 0
   \\[0.4cm]
     g+\frac{E}{2g}\left [ \ln \left (\frac{-E}{4g^2} \right )-2\mathbf{C}+2\right ] 
   + \mathcal{O}(E^2\ln(-E)) 
   & \mbox{for } \mu=1
   \\[0.3cm]
   \mu g + 2\frac{\mu g \Gamma(1-\mu)}{\Gamma(1+\mu)}\Big(\frac{-E}{4g^2}\Big)^{\mu}
       + \frac{\Gamma(1-\mu)}{\Gamma(2-\mu)}\frac{E}{2g}
       + \mathcal{O}((-E)^{1+\mu}) 
   & \\
   &\hspace{-3cm}\mbox{for } 0<\mu<1 \mbox{ and } 1<\mu<2 
 \end{array}
 \right.
\end{align*}
where
$\mathbf{C}=0.577...$ is the Euler--Mascheroni constant. 
It is useful to point out that, in the case $\mu=0$, the next correction in the expansion is $\mathcal{O}( E/\ln(-E) )$.
Analytic continuation then gives
\begin{equation}
  \label{eq:SusyLowEnergyBehaviours}
  N(E)
  \underset{E\to0^+}{=}
  \left\{
    \begin{array}{ll}
      \frac{g/2}{\left [ \ln \left ( \frac{2g}{\sqrt{E}} \right )-\mathbf{C} \right ]^2+\frac{\pi^2}{4}} + \mathcal{O} \left ( \frac{E}{\ln^2E} \right ) 
        & \mbox{for } \mu = 0
        \\[0.4cm]
      \frac{2g}{\Gamma(\mu)^2}\Big(\frac{E}{4g^2}\Big)^\mu + \mathcal{O}(E^{1+\mu})
        &  \mbox{for } \mu>0
    \end{array}
  \right.
\end{equation}
and
\begin{equation}
  \label{eq:SusyLowEnergyBehaviours2}
  \hspace{-0.75cm}
  \gamma(E)
  \underset{E\to0^+}{=}
  \left\{
    \begin{array}{ll}
      g\,\frac{\ln \left ( \frac{2g}{\sqrt{E}} \right )-\mathbf{C}}{\left [ \ln \left (\frac{2g}{\sqrt{E}} \right )-\mathbf{C} \right ]^2+\frac{\pi^2}{4}} 
        + \mathcal{O}(E) 
        & \mbox{for } \mu = 0
        \\[0.4cm]
      \mu g + \frac{2\mu g \Gamma(1-\mu)}{\Gamma(1+\mu)}\cos(\pi\mu)\big(\frac{E}{4g^2}\big)^{\mu}
        +\mathcal{O}(E) 
        & \mbox{for } 0<\mu<1 
        \\[0.3cm]
      g + \frac{E}{2g}[\ln \left ( \frac{E}{4g^2} \right )-2\mathbf{C}+2]+ \mathcal{O}(E\ln^2E) 
        & \mbox{for } \mu=1
        \\[0.3cm]
      \mu g + \frac{\Gamma(1-\mu)}{\Gamma(2-\mu)}\frac{E}{2g} +\mathcal{O}(E^{\mu}) 
        & \mbox{for } 1<\mu<2
    \end{array}
   \right.
\end{equation}
For $\mu=0$, the behaviour  $N(E)\sim g/\ln^2E\to0$ is another expression of the Dyson singularity of the density of states $N'(E)\sim1/(E|\ln E|^3)$~; see
Figure~\ref{fig:ComparisonSusyNieuwenhuizen}.
The behaviour $\gamma(E)\sim g/|\ln E|\to0$, depicted  in Figure~\ref{fig:ComparisonSusyNieuwenhuizen}, implies the divergence of the localisation length  $1/\gamma$. This fact is related to the existence of a critical state at $E=0$ characterized by power law correlations;
see \cite{SheTsv98,ComTex98}.
For a critical discussion of the interpretation of $1/\gamma$ as the localisation length in the case $\mphi=0$, see \cite{TexHag10,ComTexTou13}.

\subsection{Scalar noise with exponentially-distributed weights}
\label{sec:LimitBehaviourScalar}

It will be convenient to  express the characteristic function (\ref{eq:ResYves}) in terms of the Kummer  (confluent hypergeometric) function~\cite{gragra}~:
\begin{equation}
  \label{eq:ScalarImpurities}
    \Omega(k^2+\I0^+)
    = - \I k - \iw + (\rho-2\I k)
    \frac{ \Psi\big(2+\frac{\I\rho}{2k},3,-2\I k/\iw\big) }
    { \Psi\big(1+\frac{\I\rho}{2k},2,-2\I k/\iw\big) }
    \:.
\end{equation}
The first term on the right-hand side is the exponent obtained in the free case, i.e. in the absence of disorder.
We now examine various limits.

\paragraph{High density of scatterers $\rho\gg\iw$.---}

\begin{figure}[htbp]
  \centering
  \includegraphics[width=0.475\textwidth]{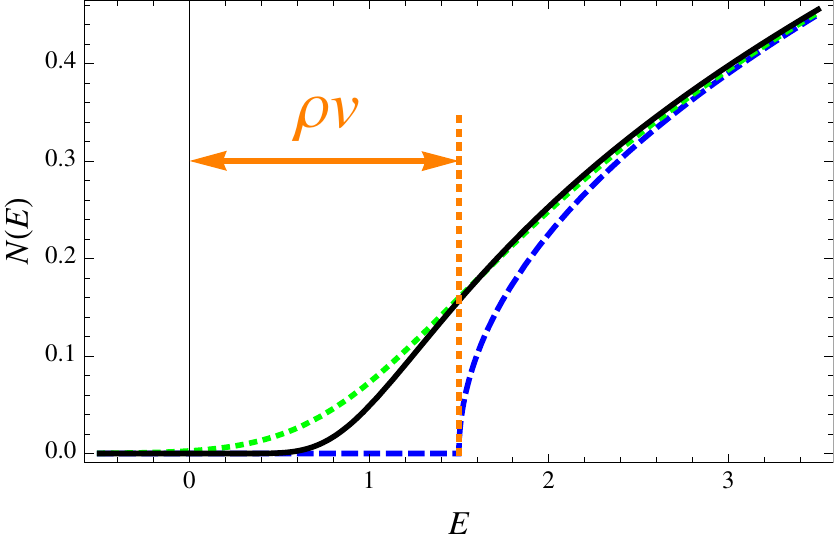}
  \hspace{0.25cm}
  \includegraphics[width=0.475\textwidth]{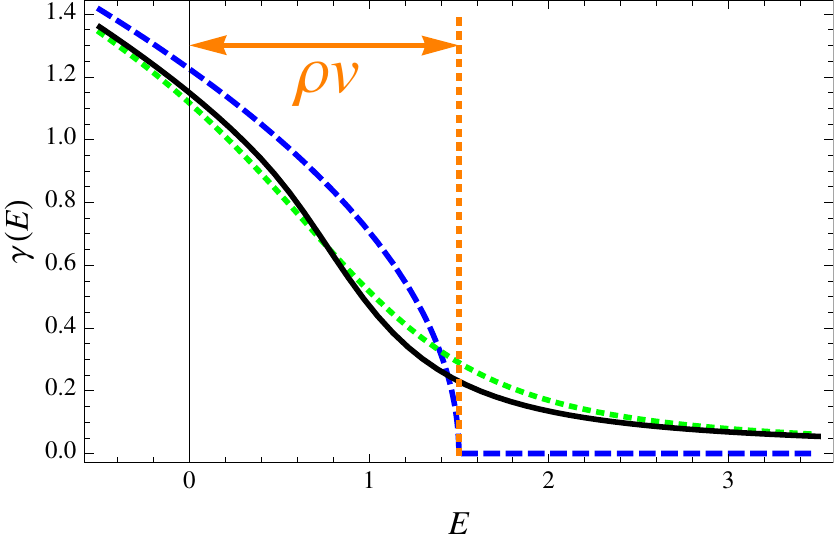}
  \caption{\it 
    Integrated density of states and Lyapunov exponent for monolithic scalar disorder
    with a high density of impurities and small positive weights. The black solid curves are obtained by using Formula \eqref{eq:ScalarImpurities}
    with $\rho=4.5$ and $v=1/3$. The
    blue dashed curves correspond to the free problem with a shift
    $\rho v$ of the energy. The
    green dotted curves correspond to the results obtained 
    for a Gaussian white noise of variance $\sigma=2\rho v^2=1$ with a shift 
    $\rho v$ of the energy and are obtained from Eq.~\eqref{eq:Halperin}.}  
  \label{fig:FLpwhd}
\end{figure}

In the high density limit, up to a shift $\rho\,\smean{v_n}=\rho v$ of the energy, the random
potential $\Absor(x)$ approximates a Gaussian white noise of covariance
$$
\mathrm{cov}[\Absor(x),\Absor(x')]=\sigma\,\delta(x-x') \;\;\text{with} \;\sigma=\rho\,\smean{v_n^2}=2\rho v^2\,.
$$
The density of states and the Lyapunov exponent are well known in this case \cite{Hal65,LifGrePas88} and may be expressed in terms
of the Airy functions:
\begin{equation}
  \label{eq:Halperin}
  \Omega(E+\I0^+)
  =\left(\frac{\sigma}{2}\right)^{1/3}
  \frac{\mathrm{Ai}'(z)-\I\,\mathrm{Bi}'(z)}{\mathrm{Ai}(z)-\I\,\mathrm{Bi}(z)}
  \hspace{0.5cm}
\text{where} \;\;
  z=-\left(\frac{2}{\sigma}\right)^{2/3}E
   \:.
\end{equation}
One may easily verify that \eqref{eq:ScalarImpurities} does indeed yield \eqref{eq:Halperin} as $\rho\to\infty$ and $v\to0$ with $\sigma=2\rho v^2$ fixed~; 
results from the two models are shown in Figure~\ref{fig:FLpwhd} for $\rho=4.5$ and $v=1/3$. We see that \text{the} two curves agree more closely
for the higher density $\rho$.

\begin{figure}[htbp]
  \centering
  \includegraphics[width=0.475\textwidth]{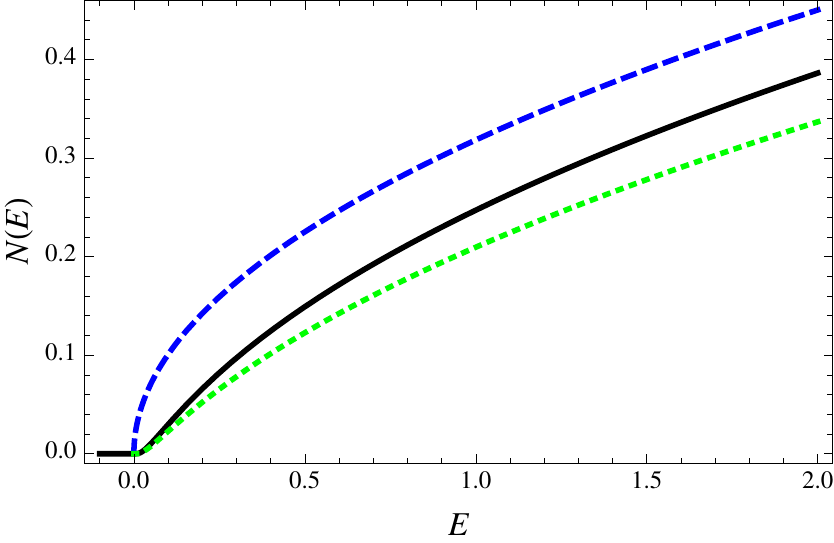}
  \hfill
  \includegraphics[width=0.475\textwidth]{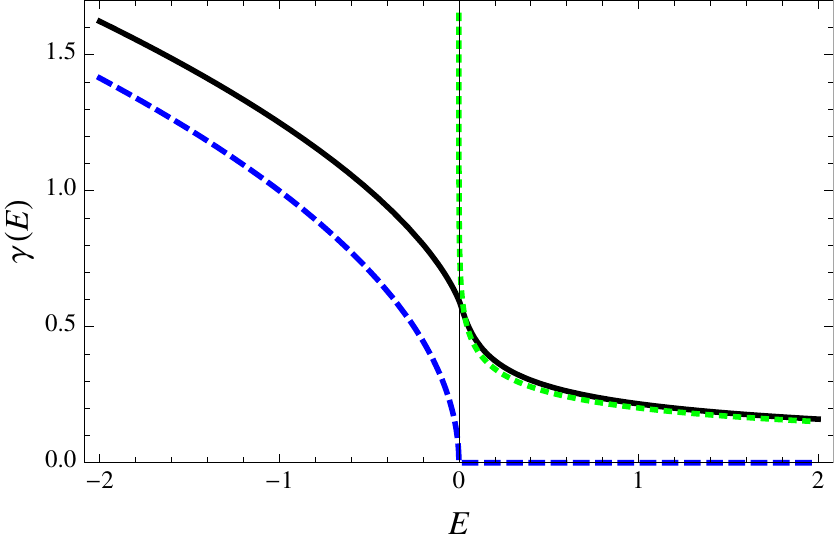}
  \caption{\it The integrated 
    density of states and the Lyapunov exponent for monolithic scalar
    disorder, as given by Eq.~(\ref{eq:ScalarImpurities}),  for a low density of impurities and large
    positive weights 
    ($\rho =0.25$ and $\iw=5$). 
    The blue dashed curve corresponds to the free problem.
    The green dotted curves are approximations discussed in
    the text.} 
  \label{fig:FLpwld}
\end{figure}

\paragraph{Low density of scatterers $\rho\ll\iw$.---}
The density of states exhibits the well-known Lifshits singularity at $E=0$, which is explained by the presence of low-lying eigenstates trapped in large regions free of impurities \cite{LifGrePas88}.
The heuristic argument first proposed by Lifshits goes like this~:
for a low energy and large $v_n$, the effect of the delta peaks in the potential is the same as if 
a Dirichlet
condition were imposed at each impurity $x_n$. The Dirichlet problems--- one for each subinterval $[x_n,x_{n+1}]$--- are completely decoupled.
The probability that
a state of low energy $E_1=(\pi/\ell)^2$ occurs is
therefore related to the probability of having an interval of large length
$\ell$ without impurity~:
\begin{equation}
  \label{eq:Lifshits}
  N(E) \underset{E\to0}{\simeq} 
  \rho\,\proba[E_1<E]
  =\rho\,\proba[\ell>\pi/\sqrt{E}]=\rho\,\e^{-\pi\rho/\sqrt{E}}
 \:.
\end{equation}
A better approximation consists of adding the contributions
from all the intervals;
this is known as the \og pieces model \fg{} \cite{GreMolSud83} and yields
$$
N(E)\simeq \frac{\rho}{ \e^{\pi\rho/\sqrt{E}}-1 }
\,.
$$ 
This approximation
is compared to the exact result in Figure~\ref{fig:FLpwld}. 

Turning next to the Lyapunov exponent, a good approximation may we obtained by
using the so-called \og concentration expansion \fg{} \cite{LifGrePas88,BieTex08}~:
$$
\gamma(k^2) \simeq \frac{\rho}{2}\,{\mathbb E}  \left( \ln \left [ 1+\left ( \frac{v_n}{2k} \right )^2 
\right ]  \right)\,.
$$
The result of averaging over the weights $v_n$ is shown in Figure~\ref{fig:FLpwld} (green dotted line).
In the high energy limit ($\rho\ll v\ll k$) we find the perturbative
result
$$
\gamma(k^2)\simeq{\rho v^2}/{(2k)^2}
$$
while in the intermediate regime  ($\rho\ll k\ll v$) we find a
logarithmic behaviour 
$$
\gamma(k^2)\simeq\rho\ln(v/2k)\,.
$$

As $E \rightarrow 0^+$, the Lyapunov exponent tends to a strictly positive limit  that may be found
by using the representation (\ref{eq:ScalarImpurities}). For the denominator, we write
$$
  \Psi\big(1+\frac{\I\rho}{2k},2,-2\I k/\iw\big)
  = \frac{\I\,\iw}{2k\,\Gamma(1+\frac{\I\rho}{2k})}
  \underbrace{
    \int_0^\infty\D t\,\e^{-t}\,
    \left(1-\frac{2\I k}{\iw}\frac1t\right)^{-\frac{\I\rho}{2k}}
  }_{
    \underset{k\to0}{\longrightarrow}
    \sqrt{\rho/\iw}\,K_1\left(2\sqrt{\rho/\iw}\right)
  }
  \:.
$$
Proceeding in a similar way for the numerator, we get
\begin{eqnarray}
  \Omega(0) 
  =\gamma(0)
  = \iw
  \left( -1 + \sqrt{\rho/\iw}\,
    \frac{K_2\big(2\sqrt{\rho/\iw}\,\big)}{K_1\big(2\sqrt{\rho/\iw}\,\big)}
  \right)\,.
\end{eqnarray}
This form is reminiscent of the result (\ref{eq:LyapsusyNegativeEnergy}) for $\mu=1$.

In the limit of a high density of impurities ($\rho\gg\iw$), we obtain
$$
\gamma(0)\simeq\sqrt{\rho\iw}\,.
$$
In words, the zero-energy Lyapunov exponent is the square root of the noise's mean value.
This result simply coincides with the result for the free
Hamiltonian, $\gamma(E)=\sqrt{-E}$ after a shift 
$\rho\iw$ of the energy~; see Figure~\ref{fig:FLpwhd}. 

The limit of low density ($\rho\ll\iw$) may by studied by using the small-argument asymptotics of the MacDonald 
function. We find
\begin{equation}
   \gamma(0)\simeq\rho\, \left [ \ln(v/\rho)-2\mathbf{C} \right ] 
  \:.
\end{equation}
This result should be compared with that obtained when the weights are fixed and all equal to $v$,
namely
\begin{equation}
  \label{eq:LyapunovOFixedWeights}
 \gamma(0)\simeq\rho\,\left [ \ln(v/\rho)-\mathbf{C}\right ] 
 \:.
\end{equation} 
We see that the two expressions differ only in the subdominant term.
Gredeskul {\it et al.} in \cite{LifGrePas88}, \S~10.4, pointed out
the logarithmic behaviour $\gamma(0)\sim-\rho\ln(\rho)$.


\section{Mixed case with fixed weights}
\label{sec:MixedCaseFixedWeights}

We begin our study of the mixed case with a particular model that
illustrates well the difficulty of extracting analytical results from the Dyson--Schmidt approach~: that where
$\IntAbsor(x)$ is a Poisson process.
The distribution of the weights $v_n$ in Equation (\ref{absorptionProcess}) is then
\begin{equation}
  p(y)=\delta(y-v)
  \:.
\end{equation}
Equivalently,
\begin{equation}
  \label{fixedWeightExponent}
  \Lambda(q)=\rho\,(1-\e^{-\I qv})\,.
\end{equation}
This model was first analysed by Texier \& Hagendorf  in \cite{TexHag09}, and later by Le Doussal \cite{LeD09}~; both studies were confined to  the limit $v\to\infty$ (cf. Figure~\ref{fig:PhaseDiagram}). 

We do not know how to solve our fundamental equation (\ref{eq:FdeQ}) in this case. We shall instead work directly
with the integro-differential equation (\ref{eq:IntegralEquationForF}) for the stationary density $f$, and confine our attention
to the special case where $\mu=E=0$. The restriction on $\mu$ is merely a matter of convenience, but that on $E$ seems unavoidable (see the remark at the end of Section~\ref{sec:MixedCaseFixedWeights}). 
The equation for $f$ then reads 
\begin{equation}
  \label{eq:Tat0}
  0 = z^2f(z) + 2gz\deriv{}{z}[zf(z)] - \rho\int_{z-v}^z\D t\,f(t) 
  \:.
\end{equation}
We proceed to show how $f$ may in principle be computed explicitly in the interval $(0\,,\,n v)$ by recurrence on the natural number~$n$.

\subsection{The interval $0< z < v$}

The key observation is that
$f(z)$ is supported on $\mathbb{R}_+$.
The lower limit of integration in the integral on the right-hand side of 
Eq.~(\ref{eq:Tat0}) may therefore be replaced by  $0$. The
equation then takes the form of a simple differential equation for the new unknown
\begin{equation}
  \label{eq:7}
  \int_{0}^z\D t\,f(t)  =:
  \frac{\normeF }{\sqrt{\xi}}\e^{-\frac\xi2}\,
  F_1(\xi)
  \hspace{0.25cm}\mbox{ with }\hspace{0.25cm} \xi := z/2g
\end{equation}
where $\normeF$ is a normalisation constant. $F_1$ solves the Whittaker equation~\cite{gragra}
\begin{equation}
  \label{eq:diffPhi1}
  \left[\deriv{^2}{\xi^2}
    - \frac14 - \frac1{2\xi} 
    +\left(\frac14-\frac\rho{2g}\right) \frac1{\xi^2}
  \right] F_1(\xi) = 0
\end{equation}
As a check, we see that, when $\rho=0$, the differential equation admits the solution 
  \begin{equation}
    \label{eq:8}
    F_1(\xi) = \sqrt{\xi}\,\e^{\frac{1}{2}\xi}
    \hspace{0.5cm} \text{i.e.} \hspace{0.5cm} f(z)=\delta(z)
  \end{equation}
as is clearly expected from Equation~(\ref{eq:RicEq}).
The general solution of (\ref{eq:7}) may be written as a linear combination of the Whittaker functions 
$M_{-1/2,\nu/2}(\xi)$ and $M_{-1/2,-\nu/2}(\xi)$ where
$$
  \nu = \sqrt{2\rho/g}
  \:.
$$
The solution must be regular at $\xi=0$. Hence
$$
  F_1(\xi) = M_{-1/2,\nu/2}(\xi)
$$
From the asymptotics of the Whittaker function~\footnote{
  \label{footnote:Useful1}
  We have $M_{\lambda,\mu}(z) \underset{z\to0}{\simeq} z^{\mu+1/2}$ and 
  $$
  M_{\lambda,\mu}(z) \underset{z\to+\infty}{\simeq}
  \frac{\Gamma(2\mu+1)}{\Gamma(\frac12+\mu-\lambda)}
  \frac{\e^{z/2}}{z^\lambda}
  \sum_{n=0}^\infty\left(\frac12+\lambda-\mu\right)_n\left(\frac12+\lambda+\mu\right)_n\frac{1}{n!z^n}\,.
  $$
This last representation can be found in \cite{DLMF}, Formula (13.19.1).
}
we deduce the limiting behaviours
\begin{equation}
  \label{eq:15}
  \int_{0}^z\D z'\,f(z') \simeq \normeF \times
  \left\{
    \begin{array}{lll}
      {\displaystyle  \left(\frac{z}{2g}\right)^{\nu/2}  }
      &\mbox{ for } & z\ll g \\[0.5cm]
      {\displaystyle  
        \frac{\Gamma(\nu+1)}{\Gamma(\nu/2+1)}
        \left(1-\frac\rho{z}+\mathcal{O}(z^{-2})\right)
      }
      &\mbox{ for } &  g \ll z < v
    \end{array}
  \right.
\end{equation}
This significanlty improves on the approximate solution obtained by
qualitative arguments in Ref.~\cite{TexHag09}. 

Equation (13.15.20) of \cite{DLMF} says
\begin{equation}
  \label{eq:Useful3}
  \deriv{}{\xi}\left[
    \frac{\e^{-\xi/2}}{\sqrt{\xi}}M_{-1/2,\nu/2}(\xi) 
  \right] = 
  \frac{\nu}{2\xi}\,\frac{\e^{-\xi/2}}{\sqrt{\xi}}M_{+1/2,\nu/2}(\xi) 
  \:.
\end{equation}
We deduce, for $0 <  z < v$,
\begin{multline}
  \label{eq:fOnFirstInterval}
    f(z) = \normeF \,
    \sqrt{\rho}\: \frac{\e^{-\frac{z}{4g}}}{z^{3/2}}\,
    M_{\frac12,\frac\nu2}\left(\frac{z}{2g}\right) \\
    = \frac{\normeF \,\sqrt{\rho}}{ (2g)^{\frac{5-\nu}2}}\,
    z^{-1+\frac\nu2}\e^{-\frac{z}{2g}}\,
    \Phi\left( \frac\nu2 , 1+\nu ; \frac{z}{2g} \right)
\end{multline}
where $\Phi(a,c;z)$ is a Kummer function~\cite{gragra}.
This is plotted for various values of $\rho$ in Figure~\ref{fig:DisRic}.

\subsection{The interval $v< z < 2v$}

\begin{figure}[!ht]
  \centering
  \includegraphics[width=0.6\textwidth]{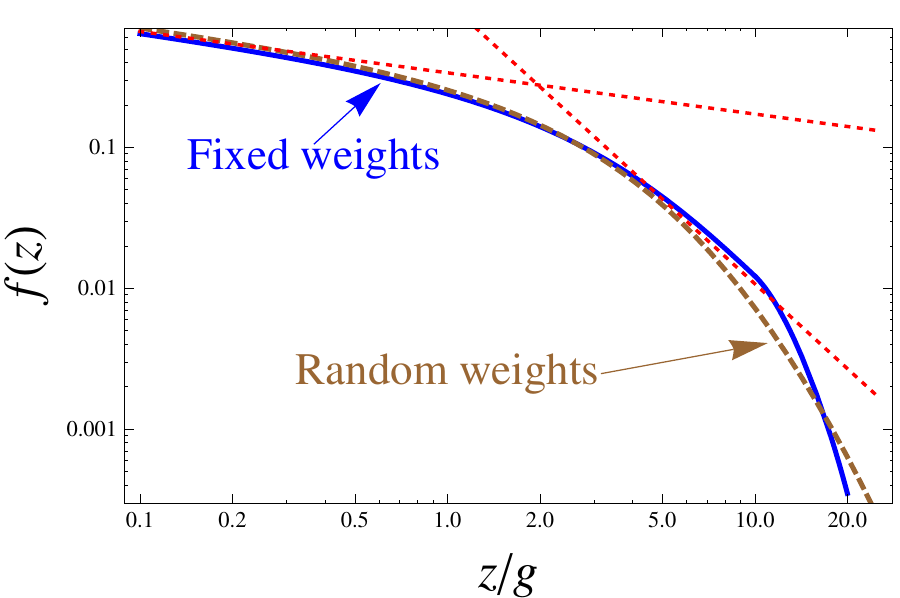}
  \caption{\it Distribution of the Riccati variable for $E=0$, $g=1$, $\rho=1$ and $v=10$ (i.e. $\nu=\sqrt{2}$). The two dotted lines correspond to the asymptotics of expression (\ref{eq:fOnFirstInterval})~: $z^{\nu/2-1}$ and $z^{-2}$. 
For comparison, we have also superimposed a plot  (dashed line) of the distribution corresponding to exponentially-distributed weights~; see Eq.~(\ref{eq:ZeroEnergyFdeZ}).} 
  \label{fig:Distf0}
\end{figure}

We now use $\int_{z-v}^z=\int_{z-v}^v+\int_{v}^z$ in (\ref{eq:Tat0}) to obtain
\begin{equation}
  \label{eq:Tat0bis}
  z^2f(z) +2gz\deriv{}{z}[zf(z)] 
  - \rho\int_{v}^z\D t\,f(t)
  =  
  \rho\int_{z-v}^v\D t\, f(t)
\end{equation}
By virtue of our earlier calculation, the right-hand side is known. Set
\begin{equation}
  \int_{v}^z\D t\,f(t)  =:
  \frac{\normeF }{\sqrt{\xi}}\e^{-\frac\xi2}\,
  F_2(\xi)
  \:.
\end{equation}
The new unknown $F_2$ solves
\begin{equation}
  \label{eq:diffPhi2}
  \left[\deriv{^2}{\xi^2}
    - \frac14 - \frac1{2\xi} 
    +\frac{1-\nu^2}{4} \frac1{\xi^2}
  \right] F_2(\xi) = 
  \nu\,R_1(\xi)
  \:,
\end{equation}
where
\begin{multline}
  R_1(\xi) :=
  \frac{\nu}{4}\frac{F_1(v/2g)-F_1(\xi-v/2g)}{\xi^2} \\
  =\frac{\nu}{4\xi^2}
  \left[
    M_{-\frac12,\frac\nu2}(v/2g)-M_{-\frac12,\frac\nu2}(\xi-v/2g)
  \right]
  \:.
\end{multline}
The particular solution we seek must vanish at $v/2g$. Furthermore, the requirement that $f$ should be continuous at $v$ translates into the condition
$$
  \deriv{}{\xi}\left[
    \frac{\e^{-\xi/2}}{\sqrt{\xi}}F_2(\xi) 
  \right]_{\xi=v/2g} = 
  \deriv{}{\xi}\left[
    \frac{\e^{-\xi/2}}{\sqrt{\xi}}F_1(\xi) 
  \right]_{\xi=v/2g} \,.
$$
By using the method of variation of constants, we eventually find the following formula, valid
for $v < z < 2 v$:
\begin{eqnarray}
  \label{eq:fOnSecondInterval}
  &&f(z) = \normeF \,
  \sqrt{\rho}\: \frac{\e^{-z/4g}}{z^{3/2}}\,
   \\\nonumber
  && \times\bigg\{
  \frac g{v}M_{\frac12,\frac\nu2}(v/2g)
  \Big[ 
       M_{-\frac12,-\frac\nu2}(v/2g)\,M_{\frac12,\frac\nu2}(\xi) 
   +  M_{-\frac12,\frac\nu2}(v/2g)\,M_{\frac12,-\frac\nu2}(\xi)
  \Big]
   \\\nonumber
  &&+
      M_{\frac12,\frac\nu2}(\xi)\int_{v/2g}^\xi\D\eta
    \,R_1(\eta)\,M_{-\frac12,-\frac\nu2}(\eta)
    + M_{\frac12,-\frac\nu2}(\xi)\int_{v/2g}^\xi\D\eta
    \,R_1(\eta)\,M_{-\frac12,\frac\nu2}(\eta)
  \bigg\}
  \:.
\end{eqnarray}
A partial plot of $f$ based on the formulae (\ref{eq:fOnFirstInterval}) and (\ref{eq:fOnSecondInterval}) is shown in Figure~\ref{fig:Distf0}.
The normalisation constant $C$ is not computed exactly~; rather, it is approximated by
setting $f$ to zero for $z>2v$. This approximation becomes exact in the limit $v \rightarrow \infty$.
The discontinuity of the slope at $z=v$ is a consequence of the fact that the absorption process has jumps of a fixed height~$v$.

\subsection{The intervals $(n-1)v< z <nv$}
The inductive nature of the approach is now clear~: given the
solution $f(z)$ on the interval $(n-2)v<z<(n-1)v$ for some natural number $n$, we can compute $f(z)$ on the next interval as follows:
we introduce the new unknown
\begin{equation}
  \int_{(n-1)v}^z\D t\,f(t)  =:
  \frac{\normeF }{\sqrt{\xi}}\e^{-\frac\xi2}\,
  F_n(\xi)
  \hspace{0.5cm} \mbox{ for } z \in [(n-1)v,nv]
  \:.
\end{equation}
$F_n$ solves an equation with the same homogeneous part as Equation  (\ref{eq:diffPhi2}), but the right-hand side now
involves
\begin{equation}
  R_{n-1}(\xi) := 
  \frac\nu4\frac{F_{n-1}((n-1)v/2g)-F_{n-1}(\xi-(n-1)v/2g)}{\xi^2}
  \:.
\end{equation}
However the calculations become tedious for $n > 2$.
The normalisation constant is given by 
\begin{equation}
  1/\normeF 
  = \sum_{n=1}^\infty\frac{\e^{-n v/4g}}{\sqrt{n v/2g}} F_n(nv/2g)
\end{equation}
As remarked earlier, when $v$ is large, good approximations are obtained by retaining only the first few terms.

\begin{figure}[!ht]
  \centering
  \includegraphics[width=0.6\textwidth]{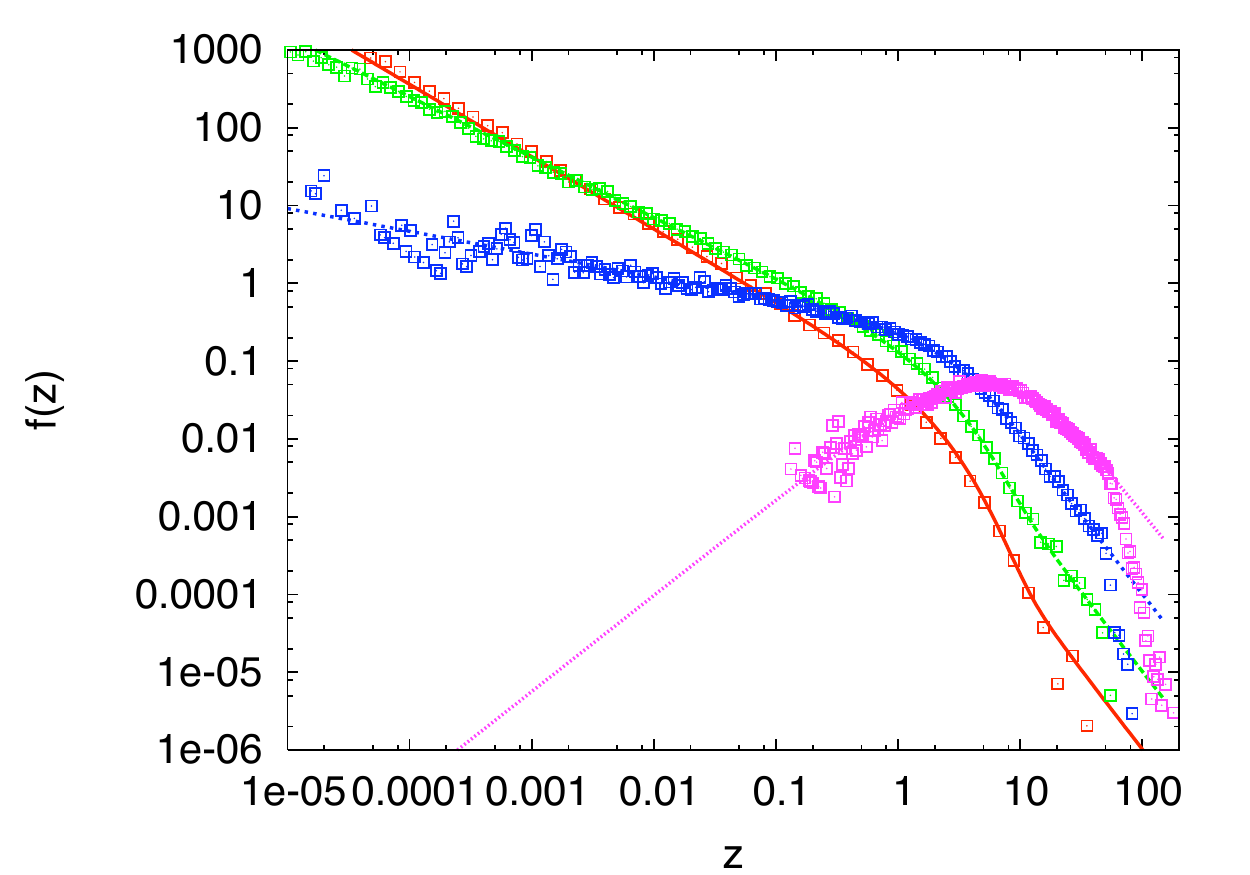}
  \caption{
    {\it The probability density $f(z)$ of the Riccati variable for
      $E=0$,  $v=50$, $g=1$ and
      various values of $\rho$, namely $\rho=0.01$, $0.1$, $1$ and $10$. 
      The solid lines are obtained by using Formula (\ref{eq:fOnFirstInterval}).
      The superimposed symbols are obtained by direct numerical simulations of the Riccati process.
    }
    }
  \label{fig:DisRic}
\end{figure}

\subsection{Numerical simulations}

The statistical properties of the stochastic Riccati equation (\ref{eq:RicEq}) may be studied by direct numerical simulation~; see
Ref.~\cite{BieTex08} for the implemention details. 
In Figure~\ref{fig:DisRic}, the results obtained from Formula (\ref{eq:fOnFirstInterval}) are compared to the results of Monte Carlo simulations (the only fitting parameter is the global multiplicative constant $C$)~; the agreement is excellent.

\subsection{Low density $\rho\to0$ and large weights $v\to\infty$}
\label{sec:FI0Lyap}

\begin{figure}[!ht]
  \centering
  \includegraphics[width=0.55\textwidth]{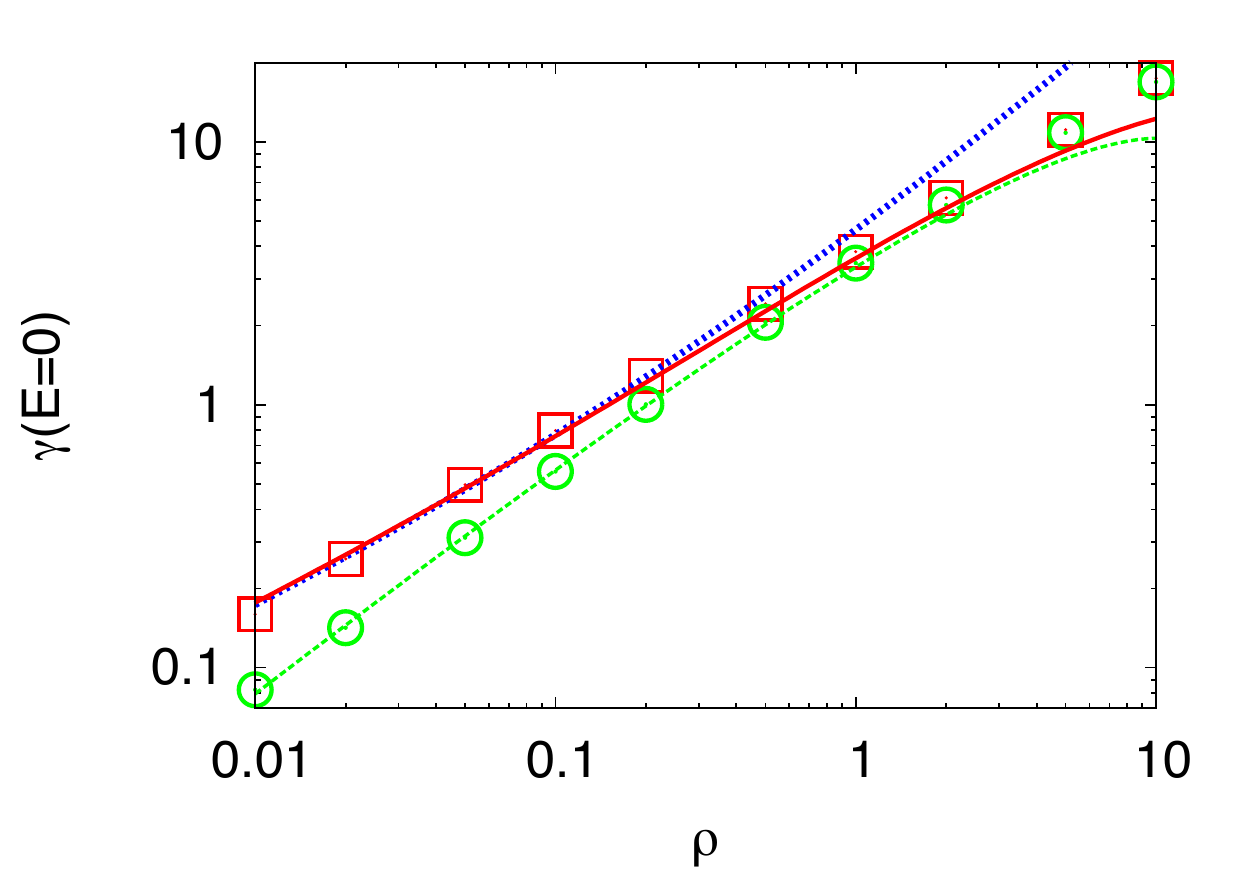}
  \caption{
    {\it The Lyapunov exponent at $E=0$ as a function of the density $\rho$ of impurities:
      the results of numerical simulations (red squares) for $g=1$ and $v=50$ are compared with $\int_0^\infty\D z\,z\,f(z)$ computed using
      (\ref{eq:fOnFirstInterval}) (red continuous line) and with the approximation 
      (\ref{eq:Lyap0v2012}) (blue dotted line).
      The green circles correspond to values obtained in the absence of the supersymmetric noise, $g=0$. This is
      compared with the analytical Formula (\ref{eq:LyapunovOFixedWeights}) (green dotted line).   
    }
  }
  \label{fig:ZeroEnergyLyapunov}
\end{figure}

Let us now discuss the limit $\rho\ll g\ll v$, corresponding to the case of interest in \cite{TexHag09,LeD09}.
We see from \eqref{eq:15} that the normalization constant is $\normeF^{-1}\simeq\frac{\Gamma(\nu+1)}{\Gamma(\nu/2+1)}
\simeq1$.
The asymptotic behaviour of the distribution is $f(z)\simeq\rho/z^2$. This may be explained qualitatively as follows 
in terms of the ``Riccati particle''
analogy mentioned earlier~: 
the process is driven towards $z=0$ by the drift (the deterministic force corresponding to the term $-z^2$) while experiencing the effect of the multiplicative noise $\phi(x)$. Whenever it meets an impurity at, say $x_n$--- an event that occurs at a rate $\rho$--- the 
Riccati particle is sent back to infinity, $z(x_n+)=z(x_n-)+v\to\infty$, where the effect of the drift is dominant and drives the particle towards $z=0$ again. 
This generates a current $\rho$ accross $\mathbb{R}_+$. (Because the deterministic force dominates at infinity, the probability density $f(z)$ is effectively
given by the ratio $\mathrm{current}/\mathrm{force}$.)
This approximation for $f$  yields, via Eq.~\eqref{eq:15}, the following estimate for the Lyapunov exponent~:
\begin{equation}
  \gamma(0) \simeq
  \frac{\nu\,\Gamma(1+\nu/2)}{2\,\Gamma(1+\nu)}
  \int_0^{2g} \D z\, \left(\frac{z}{2g}\right)^{\nu/2}
  + \rho \int_{2g}^v\frac{\D z}{z}
  \:.
\end{equation}
Finally, since $\nu\to0$, we find
\begin{equation}
  \label{eq:Lyap0v2012}
  \gamma(0) \simeq \sqrt{2\rho g} + \rho\, \ln(v/2g)
  \:.
\end{equation}
This calculation corrects (by a factor of $\sqrt{2}$) an earlier result of Ref.~\cite{TexHag09}.
The approximate expressions of the zero energy Lyapunov exponent as a function of the density of scatterers
are plotted in Figure~\ref{fig:ZeroEnergyLyapunov}, and compared with the results of Monte Carlo simulations. A comparison with the case $\phi=0$ is also made.

\begin{remark}
To end this section, let us comment briefly on the possible extension of this approach to other values of the energy.
As mentioned at the begining, the key requirement is that the stationary probability density be supported
on the positive half-line. This requirement is met when $E < 0$, but the resulting equation on the first interval $0<z<v$ belongs
to the Heun class \cite{Ron95}, and we do not know how to solve it.
\label{fixedWeightRemark}
\end{remark}


\section{Mixed case with exponentially-distributed weights~: Exact results}
\label{sec:SolvableMixedCase}

We now turn to the analysis of the differential equation (\ref{eq:FdeQ}) when $\IntAbsor(x)$ is a compound Poisson process with exponentially-distributed weights $v_n$~:
\begin{equation}
  \label{eq:LevyExpWeights}
    p(y)= \frac{1}{\iw}\e^{-y/\iw} 
    \;\;\text{for } y\geq0
    \hspace{0.5cm}\mbox{i.e. }
  \Lambda(q)= \rho\frac{\I q}{1/\iw + \I q}
\end{equation}
where $\mean{v_n}=v$ is the mean value of the weights.


Our basic equation~(\ref{eq:FdeQ}) becomes
\begin{equation}
  \label{eq:EqDiffMixedCase}
  \left[
    -(1+2\I g\,q)\deriv{^2}{q^2} + 2\I g (\mu-1)\deriv{}{q}
    +E - \frac{2g\,\rho}{2g/\iw+2\I g\,q}
  \right] \hat{f}(q) = 0  
  \:.
\end{equation}
This is a confluent limit of the 
{\it Heun differential equation}. An account of what is known about the solutions of such equations may be found
in Part B of the monograph \cite{Ron95}, but it is not easy to extract useful asymptotic estimates from this material. We shall therefore begin
by considering two particular cases where it reduces to the hypergeometric equation~\cite{gragra,NikOuv83}.

\subsection{The case $\iw\neq2g$ and $E=0$}
\label{eq:ZeroEnergyCase}

We make the substitution 
$$
z=(1+2\I g\,q)/(1-2g/\iw)
\;\;\text{and}\;\;
u(z)=\hat f(q)\,.
$$
The new unknown $u$ then satisfies the hypergeometric equation
$$
z(1-z)\,u''(z) + [c-(a+b+1)z]\,u'(z)-ab\,u(z)=0
$$
with 
$$
a=\frac12(\nu-\mu)\,, \;b=-\frac12(\nu+\mu) \; \text{and} \;c=(a+b+1)=1-\mu
$$ 
where
\begin{equation}
  \label{eq:DefNu}
  \boxed{ 
  \nu := \sqrt{\mu^2 + 2\rho/g} 
  }
\end{equation}
To find the solution vanishing for $z\to\infty$, we use the particular basis of solutions \cite{NikOuv83}
$$
z^{-a}\:_2F_1(a,a-c+1;a-b+1;1/z) \;\;\text{and} \;\;z^{-b}\:_2F_1(b,b-c+1;b-a+1;1/z)\,.
$$
The former clearly has the desired property, hence
\begin{equation}
  \label{eq:MixedZeroEnergyFofQ}
  \hat{f}(q) = c\, 
  (1+2\I g\,q)^{\frac{\mu-\nu}{2}}\,
  _2F_1\left(\frac{\nu+\mu}{2} , \frac{\nu-\mu}{2} ; 1+\nu ; \frac{1 - 2g/\iw}{1+2\I g\,q} \right)
\end{equation}
where $c$ is a normalisation constant.
Using
$$
\deriv{}{z}\big[z^{-b}\,_2F_1(a,b;c;1/z)\big]=-b\,z^{-b-1}\,_2F_1(a,b+1;c;1/z)
\:,
$$
deduced from Formula (15.5.3) of \cite{DLMF}, we get
\begin{eqnarray}
  \label{eq:Omega0SolvableModel}
    \boxed{ 
  \gamma(0)=\mu g + (\nu-\mu)\,
      g\,\frac{ 
              _2F_1\left(\frac{\nu+\mu}{2} , 1+ \frac{\nu-\mu}{2} ; 1+\nu ; 1 - 2g/\iw \right)
            }{
            _2F_1\left(\frac{\nu+\mu}{2} , \frac{\nu-\mu}{2} ; 1+\nu ; 1 - 2g/\iw \right)
            } 
   }
\end{eqnarray}

In particular, we can examine what happens when $\mu=0$, for a low density of impurities and large $v$.
Using (\ref{eq:RatioHFresult}), we find  
\begin{equation}
  \label{eq:EW0Lyap}
  \gamma(0) \simeq \sqrt{2\rho g} + \rho\,\ln(v/2g)
  \hspace{0.5cm}\mbox{for } \rho\to0\ \mbox{and}\ \iw\to\infty
  \:.
\end{equation}
This result coincides precisely with what we found earlier  for \textit{fixed} weights~; see
Equation (\ref{eq:Lyap0v2012}).
This confirms that the details of the distribution of the weights are not important in the limit~$\iw\to\infty$.

Before leaving this particular case, let us remark that, for exponentially-distributed weights, the integro-differential equation 
(\ref{eq:IntegralEquationForF}) for $f$
can be reduced to a purely differential form. This differential equation may be solved when $E=0$, and one finds
\begin{equation}
  \label{eq:ZeroEnergyFdeZ}
  f(z) = C \, z^{-1+\frac{\nu-\mu}{2}}\, e^{-z/2g}\,
  \Phi\left( \frac{\mu+\nu}{2} , 1+\nu ; \left [ 1/(2g) - 1/\iw \right ] z \right)
  \:.
\end{equation}
By using Formula (7.621.4) of Ref.~\cite{gragra}, one may verify that this is consistent
with  (\ref{eq:MixedZeroEnergyFofQ}). 
Interestingly, this probability density has a form that is very similar to the one obtained earlier for fixed weights~; compare with Eq.~(\ref{eq:fOnFirstInterval}). The two densities are shown in Figure~\ref{fig:Distf0}.

\subsection{The case $\iw=2g$}
\label{sec:SolvableCase}

This case is more informative, since we will now obtain an exact expression of the complex Lyapunov exponent for the full spectrum of energy.
Set
\begin{equation}
  x := \frac{k}{g} \sqrt{1+ 2 \I g q}
  \hspace{0.5cm}\text{and}\hspace{0.5cm}
  \hat{f}(q) =: x^\mu\, y(x)\,.
  \label{substitution}
\end{equation}
Then Equation \eqref{eq:EqDiffMixedCase} becomes
\begin{equation}
  \label{eq:5.9}
  y''(x) + \frac{1}{x} y'(x) + \left [ 1 - \frac{\nu^2}{x^2} \right ] y(x) = 0
\end{equation}
where $\nu$ is the parameter defined in Equation (\ref{eq:DefNu}).
This is the differential equation satisfied by the Bessel functions of index $\nu$.
The requirement that $\hat{f}$ decay at infinity leads to
\begin{equation}
  \hat{f}(q) = c\, (1+2\I g\,q)^{\mu/2} \, 
  H^{(1)}_{\nu}\left(\frac{k}{g}\sqrt{1+2\I g\,q}\right)
\label{exactSolutiion}
\end{equation}
where $c$ is a normalisation constant. Hence
\begin{equation}
  \label{eq:FdeQExpWeightSolvablePoint}
    \Omega(k^2+\I0^+) = -\mu g 
    + k\, \frac{ H^{(1)}_{\nu+1}\left({k}/{g}\right) }{ H^{(1)}_\nu\left({k}/{g}\right) }
  \:.
\end{equation}

We see that this complex Lyapunov exponent has exactly the same form as that for the supersymmetric Hamiltonian \eqref{eq:Hsusy}~; see  Eq.~\eqref{eq:LyapsusyPositiveEnergy}.
It is remarkable that,  at least at the level of the self-averaging quantity  $\Omega$, the effect of adding to the random supersymmetric potential the noise $\Absor(x)$ defined by
Equation (\ref{eq:LevyExpWeights}) can be accounted for through a simple transformation of the parameter $\mu=\mean{\phi(x)}/g$:
\begin{equation}
  \label{eq:SimpleSubstitution}
  \mu\to\nu=\sqrt{\mu^2+2\rho/g}
 \:.
 \end{equation}
We refer the reader to \S \ref{supersymmetricDisorderSubsection}
for a discussion of asymptotic behaviours and for plots of
$N$ and $\gamma$ for various values of the Hankel index.

\begin{figure}[!ht]
\centering
\includegraphics[width=0.475\textwidth]{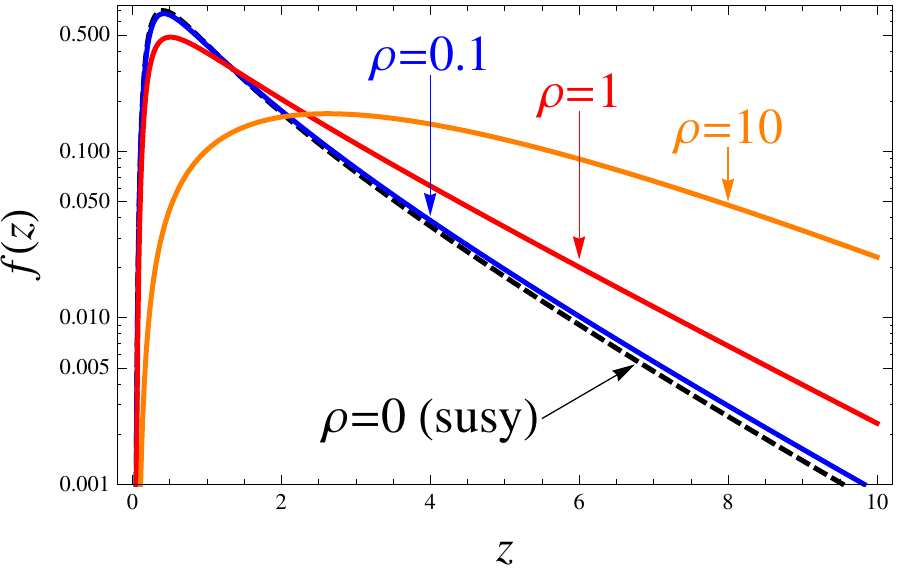}
\hfill
\includegraphics[width=0.475\textwidth]{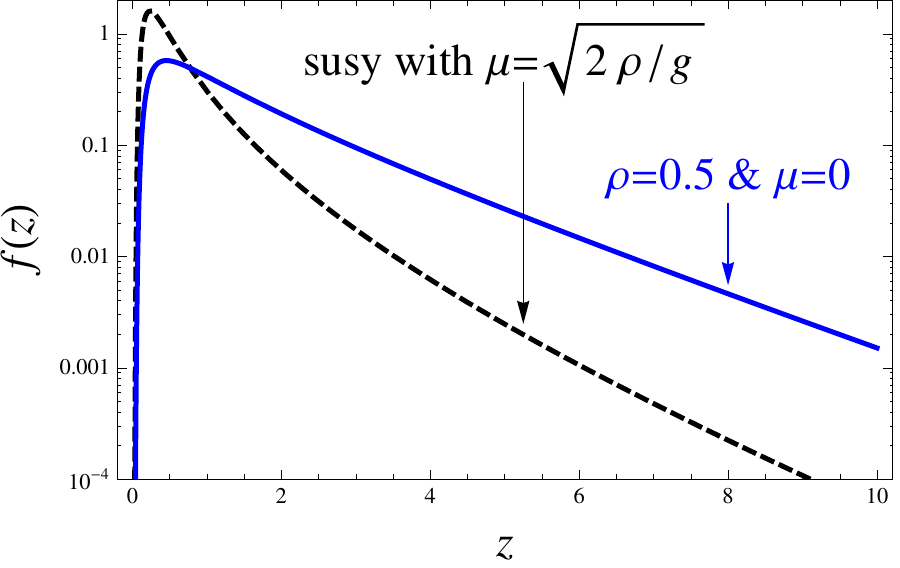}
\caption{Left~: 
         {\it The probability density (\ref{eq:DistributionAtSolvablePoint}) of the Riccati variable for $k=g=1$, $\mu=0$, and various values of $\rho$.}
         Right~:
         {\it Comparison between the the probability densities (\ref{eq:DistributionAtSolvablePoint}) for the mixed case with $\mu=0$ and $\nu=\sqrt{2\rho/g}$, and for the supersymmetric case, 
$f(z)=k^{\mu}\,z^{-1-\mu}\exp\big[-\frac{1}{2g}(z+k^2/z)\big]/\big[2K_\mu(k/g)\big]$, in which we set $\mu=\nu$.}}
\label{fig:ComparisonSusySolvable}
\end{figure}

The differential equation for the probability density $f$, which we alluded to earlier, may also be solved explicitly when $v=2g$~;
for $E=-k^2$, one finds
\begin{equation}
  \label{eq:DistributionAtSolvablePoint}
  f(z) = C \, z^{-1+\frac{\nu-\mu}{2}} \, 
  e^{-\frac{1}{2g}\left(z+k^2/z\right)}\,
  \Psi\left( -\frac{\nu+\mu}{2} , 1-\nu ; \frac{k^2}{2gz} \right)
\end{equation}
where $\Psi(a,c;z)$ is a Kummer function~\cite{gragra}.  By expressing the Kummer function in terms of a Whittaker function, making use of Formula (7.630.2) of Ref.~\cite{gragra}, and using analytic continuation in $E$, we recover (\ref{eq:FdeQExpWeightSolvablePoint}). It is interesting to note that, unlike the complex Lyapunov exponent, this probability density does not
appear to have a simple relationship to that obtained for monolithic supersymmetric disorder~; the plot on the right of Figure~\ref{fig:ComparisonSusySolvable} illustrates this point.

\subsection{Replica derivation for $\iw=2g$}
\label{subsec:Alain}

We give an alternative derivation of the simple correspondence (\ref{eq:SimpleSubstitution})  that uses the replica approach~\cite{Com10}.
The Green's function, averaged over the disorder,
$$
G(E) := {\mathbb E} \left ( \bra{x} \frac{1}{E-\mathscr{H}}  \ket{x} \right )
$$
can, by the replica trick~\cite{ItzDro89a}, be expressed in the form
\begin{equation}
  G(E)=-\frac{2}{L}\frac{\partial}{\partial E}
  \lim_{n\to0+}\frac\partial{\partial n} \mathcal{Z}_n
\end{equation}
where
\begin{multline}
\notag
  \mathcal{Z}_n =
  \mean{
  \int_{\chi(0)=\chi(L)}\mathcal{D}\chi\:
  \exp\bigg\{-\frac12\int_0^L\D x\,\chi(x)(\mathscr{H}-E)\chi(x)\bigg\}
  }
  \\
  =
  \int_{\chi(0)=\chi(L)}\mathcal{D}\chi\:
  \exp\bigg\{-\int_0^L\D x\,L(\chi(x),\Dot\chi(x))\bigg\}
\end{multline}
and the path integral runs over a $n$-component vector field $\chi(x)$
satisfying periodic boundary conditions. 
In this expression, $\Dot\chi$ denotes $\D{\chi}/\D x$. 
The contribution of the impurity potential to the action $\int\D x\,L$ is given by 
$$
  -\ln\left(\,\mean{\e^{-\frac12\int\D x\,\Absor(x)\chi(x)^2}}\,\right)
  =\frac12\int\D x\,\frac{\rho\,\chi(x)^2}{1/\iw+\frac{1}{2}\chi(x)^2}  
  \:.
$$
The details may be found in Appendix~\ref{appendix:SPLN}.
This contribution is added to the part coming from averaging over
$\phi(x)$~; see Refs.~\cite{BouComGeoLeD90} and \cite{HagTex08}. Gathering all the terms, we obtain
\begin{eqnarray}
  L(\chi,\Dot{\chi}) = 
  \frac{1}{2}\Dot{\chi}^2 
  -\frac{g\,\left(\chi\cdot\Dot{\chi}\right)^2}{2(1+ g\chi^2)}
  +\frac{\mu^2g^2\chi^2}{2(1+g\chi^2)}
  &+&\frac{\rho\,\chi^2}{2(1/\iw+\frac{1}{2}\chi^2)}-\frac{E}2\,\chi^2
  \nonumber
  \\
  \nonumber
  &+&\frac{1}{2}\delta^{(n)}(0)\,\ln\det(1+g\chi^2)
  \:.
\end{eqnarray}
We then see that, if $\iw=2g$, the potential term in the Lagrangian simplifies to
$$
  \frac{\mu^2g^2\chi^2}{2(1+g\chi^2)}
  +\frac{\rho\,\chi^2}{2(1/\iw+\frac{1}{2}\chi^2)}-\frac{E}2\,\chi^2
  =  \frac{(\mu^2+2\rho/g)g^2\chi^2}{2(1+g\chi^2)}
  -\frac{E}2\,\chi^2
  \:.
$$
This makes the substitution (\ref{eq:SimpleSubstitution}) clear.


\section{The low energy limit~: a perturbative approach}
\label{sec:HankelSeries}

As mentioned in the introduction, the low-energy limit is of considerable interest because, in terms of
the classical diffusion, the asymptotics of $N(E)$ for small $E$ immediately yield the long-time asymptotics of the probability density of
return to the starting point.
When the jumps of the absorption process are exponentially distributed with mean $\iw=2 g$, the exact solution (\ref{eq:FdeQExpWeightSolvablePoint}) yields
\begin{equation}
N(E) \simeq \frac{2 g}{\Gamma (\nu)^2} \left ( \frac{E}{4 g^2} \right )^\nu
\hspace{0.25cm} \mbox{as $E \rightarrow 0+$}\,.
\label{exactLowEnergyBehaviour}
\end{equation}
It is therefore highly desirable to understand how deviations from the solvable case influence the low-energy limit.

\subsection{Exponentially-distributed weights with $v \ne 2 g$}
\label{exponentialSubsection}

The change of variables (\ref{substitution})
maps Equation (\ref{eq:EqDiffMixedCase}) onto
\begin{equation}
\label{heunEquation}
  y''(x) + \frac{1}{x} y'(x) 
  + \left [1 -\frac{\mu^2}{x^2} - \frac{\nu^2-\mu^2}{x^2-\varepsilon}  \right ] y(x) = 0
  \:,
\end{equation}
where
\begin{equation}
  \varepsilon := \left(\frac{k}{g}\right)^2 \left( 1 - 2g/v \right)\,.
\end{equation}
Equation~(\ref{heunEquation}) is of the form considered by Figueiredo \cite{Fig05} who calls it a ``Ince limit'' of the confluent Heun equation. 
The parameter $\varepsilon$ measures the deviation from the solvable point corresponding to the Bessel equation of index $\nu$, Eq.~(\ref{eq:5.9}).

We shall again look for a solution that decays at infinity, 
in terms of which the complex Lyapunov exponent is 
\begin{equation}
  \label{complexLyapunovExponentAgain}
  \Omega (k^2) = -k \frac{y'(k/g)}{y(k/g)}
  \,.
\end{equation}
Instead of normalising the decaying solution by making use of the condition $\hat{f}(0)=\big( k/g \big )^\mu\,y(k/g)=1$, it will be more convenient for purely technical reasons to prescribe its asymptotic form at infinity:
\begin{equation}
  \label{asymptoticCondition}
  y(x) \simeq H_{\nu}^{(1)} (x) \;\;\mbox{as $x \rightarrow \infty$}
  \,.
\end{equation}
In terms of this particular solution, $N=-\im \Omega/\pi$ is given by
\begin{equation}
  \label{eq:IDoSfctY}
  N(k^2) = \frac{2g / \pi^2}{\left | y(k/g) \right |^2}\,.
\end{equation}
This formula is easily deduced by remarking that the prescription (\ref{asymptoticCondition}) implies that the Wronskian of $y$ and its conjugate takes the form
$\mathcal{W}[y^*,y]=2\I\,\im[y'y^*]=4\I/(\pi x)$.

To compute $y$, we follow Figueiredo's example and use the ansatz
\begin{equation}
  \label{hankelSeries}
  y(x) = \sum_{m \in {\mathbb Z}} y_m\, H_{\lambda+2m}^{(1)}(x)
\end{equation}
where the parameter $\lambda$ and the coefficients $y_m$ are unknowns to be determined.
To this end, we first re-arrange Equation (\ref{heunEquation}):
\begin{multline}
x^2 \left [ y''(x) + \frac{1}{x} y'(x) + \left ( 1 - \frac{\nu^2}{x^2} \right ) y(x) \right ]  \\
= \varepsilon \, \left [ y''(x) + \frac{1}{x} y'(x) + \left ( 1 - \frac{\mu^2}{x^2} \right ) y(x)
\right ] \,.
\label{eq:rearrangment}
\end{multline}
Then
\begin{multline}
  \notag
  x^2 \sum_{m \in {\mathbb Z}} y_m \left [ (\lambda- 2m)^2 -\nu^2 \right ] H_{\lambda+2 m}^{(1)} (x) \\
  = 
  \varepsilon \sum_{m \in {\mathbb Z}} y_m \left [ (\lambda- 2m)^2 -\mu^2 \right ] H_{\lambda+2 m}^{(1)} (x)\,.
\end{multline}
To proceed, we divide both sides by $x^2$ and use the identity
\begin{equation}
  \label{figueiredoRelation}
  \frac{1}{x^2} H_{\beta}^{(1)} (x) = \frac{H_{\beta-2}^{(1)}(x)}{4 \beta (\beta-1)}  + \frac{H_{\beta}^{(1)}(x)}{2 (\beta^2-1)} 
  + \frac{H_{\beta+2}^{(1)}(x)}{4 \beta (\beta+1)}\,.
\end{equation}
After equating the coefficients of the Hankel functions, we find that the differential equation (\ref{heunEquation}) for $y$ implies the following three-term recurrence relation for the $y_m$:
\begin{equation}
  \label{threeTermRecurrence}
  a_m y_{m+1} + b_m y_m + c_m y_{m-1} = 0
\end{equation}
where
\begin{equation*}
a_m := \frac{\varepsilon}{4} \frac{(\lambda + 2m +2)^2-\mu^2}{(\lambda + 2m +2) (\lambda+2m+1)}\,,
\end{equation*}
\begin{equation*}
b_m := \nu^2 - ( \lambda +2 m)^2 + \frac{\varepsilon}{2} \frac{(\lambda + 2m)^2-\mu^2}{(\lambda + 2m)^2-1}
\end{equation*}
and
\begin{equation*}
c_m := \frac{\varepsilon}{4} \frac{(\lambda + 2m -2)^2-\mu^2}{(\lambda + 2m -2) (\lambda+2m-1)}\,.
\end{equation*}
The fact that the recurrence relation for the $y_m$ is of order two
is quite remarkable here, and relies crucially on the rearrangement (\ref{eq:rearrangment}).
The asymptotic condition (\ref{asymptoticCondition}) translates into
\begin{equation}
\sum_{m \in {\mathbb Z}} (-1)^m y_m = \exp \left [ \I \frac{\pi}{2} (\lambda-\nu) \right ]\,.
\label{threeTermCondition}
\end{equation}

For $\varepsilon \ne 0$, there is no obvious solution of the three-term recurrence relation (\ref{threeTermRecurrence}). Neither is it clear how to
select the somewhat mysterious parameter $\lambda$. 
However, the form of the coefficients suggests that we look
for an expansion in ascending powers of $\varepsilon$:
\begin{equation}
  \lambda := \sum_{n=0}^\infty \lambda_n \, \varepsilon^n 
  \;\;\mbox{and}\;\; 
  y_m := \sum_{n=0}^\infty y_{m,n} \, \varepsilon^n
  \,.
  \label{seriesSolution}
\end{equation}
The decision to make the index $\lambda$ depend on the perturbation parameter may surprise at first sight. We are guided by the fact that our problem
has close connections with that of studying the low-energy scattering of waves by a central long-range potential, where expansions
of the form (\ref{seriesSolution}) have been found useful \cite{SadBohCavEsrFabMacRau00}. 
We refer the reader to Cavagnero \cite{Cav94}
for a clear, physically-motivated explanation
of why, when seeking to determine the series
(\ref{hankelSeries}) perturbatively,
 it is necessary to expand the index.

In order to determine the coefficients $\lambda_n$ and $y_{m,n}$ we set
$$
a_m := \sum_{n=0}^\infty a_{m,n} \, \varepsilon^n
$$
and similarly for $b_m$ and $c_m$. We then insert these expansions in (\ref{threeTermRecurrence}) and equate the coefficients
of like powers in $\varepsilon$.

To lowest order, we obtain
$$
a_{m,0} \,y_{m+1,0} + b_{m,0} \,y_{m,0} + c_{m,0} \,y_{m-1,0} = 0\,.
$$
Since $a_{m,0} = c_{m,0}=0$ and $b_{m,0} = \nu^2-(\lambda_0+2m)^2$, this reduces to
$$
\left [ \nu^2 - (\lambda_0+2m)^2 \right ] y_{m,0} = 0\,.
$$
To obtain a non-trivial solution, we require the coefficient in square brackets to vanish for some $m$. We may without loss
of generality make it vanish for $m=0$. Hence
$$
\lambda_0 = \nu\,.
$$
Then $y_{m,0} = 0$ for every $m \ne 0$ and the series reduces to a single term. The coefficient $y_{0,0}$ may be determined by
imposing the condition (\ref{threeTermCondition}). This gives $y_{0,0} =1$
and thus, unsurprisingly,
$$
y(x) = H_{\nu}^{(1)} (x) + {\mathcal O} (\varepsilon) \;\;\mbox{as $\varepsilon \rightarrow 0$}\,.
$$

At the next order, equating the terms of order $\varepsilon$, we obtain
\begin{equation}
b_{m,0} \,y_{m,1} = - a_{m,1} \,y_{m+1,0} - b_{m,1} \,y_{m,0} - c_{m,1} \,y_{m-1,0}\,.
\label{nextOrder}
\end{equation}
For $m=0$ this yields
$$
0 = a_{0,1} \,y_{1,0} + b_{0,1}\,y_{0,0}+c_{0,1}\,y_{-1,0} = b_{0,1} \, y_{0,0}
$$
since $b_{0,0}$, $y_{-1,0}$ and $y_{1,0}$ all vanish. The coefficient $y_{0,0}$, however, does not vanish, and so we must have
$$
0 = b_{0,1} := -2 \nu \lambda_1 + \frac{\rho/g}{\nu^2-1}\,.
$$
This determines $\lambda_1$, namely
$$
\lambda_1 = \frac{1}{2 \nu} \frac{\rho/g}{\nu^2-1}\,.
$$
For $m \ne 0$, Equation (\ref{nextOrder}) determines $y_{m,1}$~; we find
$$
y_{\pm 1,1} = \frac{\pm 1}{8 \nu} \frac{\rho/g}{(\nu \pm 1)^2}
$$
and $y_{m,1}=0$ for $| m | >1$. There remains to determine $y_{0,1}$. For this we make use of the asymptotic
condition (\ref{threeTermCondition}), which translates into
$$
- y_{-1,1} + y_{0,1} - y_{1,1} = \I \frac{\pi}{2} \lambda_1\,.
$$
The end result is
\begin{multline}
y (x) =  -\frac{\varepsilon}{8 \nu}  \frac{\rho/g}{(\nu-1)^2} \,H_{\lambda-2}^{(1)} (x)
+ \frac{\varepsilon}{8 \nu} \frac{\rho/g}{(\nu+1)^2} \,H_{\lambda+2}^{(1)} (x)
 \\
+\left [ 1 + \varepsilon \frac{\rho/g}{8 \nu (\nu^2-1)} \left ( 2 \I \pi + \frac{\nu-1}{\nu+1} - \frac{\nu+1}{\nu-1} \right ) \right ] \,H_{\lambda}^{(1)} (x) + {\mathcal O} \left ( \varepsilon^2 \right )
\label{firstOrderCorrectionForY}
\end{multline}
where
\begin{equation} 
\lambda = \nu + \frac{\varepsilon}{2 \nu} \frac{\rho/g}{\nu^2-1} + {\mathcal O} \left ( \varepsilon^2 \right ) \;\;\mbox{as $\varepsilon \rightarrow 0$}\,.
\label{firstOrderCorrectionForLambda}
\end{equation}

\begin{figure}[!ht]
 \centering
 \includegraphics[width=0.6\textwidth]{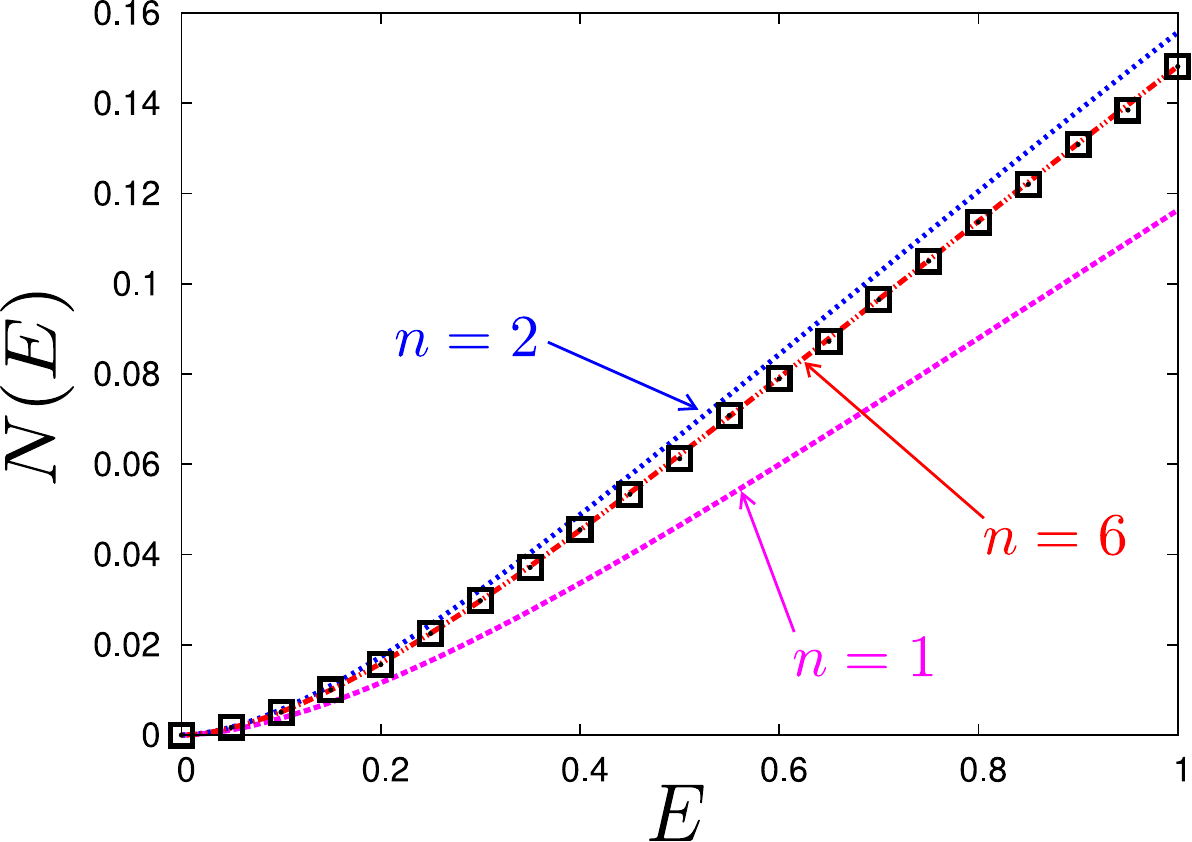}
 \caption{\it The integrated density of states at low energy for the mixed case
 with exponentially distributed weights and $\mu=g=v=1$. The squares are the results of a direct numerical simulation.
 The other curves are obtained by using ${\mathcal O}(\varepsilon^n)$ truncations of $y_m$ and $\lambda$ in the Hankel series solution (\ref{hankelSeries}) and then substituting the resulting approximation in Formula~(\ref{eq:IDoSfctY}).
 We observe rapid convergence as $n$ increases. 
}
 \label{fig:Aurelien}
\end{figure}

The procedure can be continued systematically~; for $n>1$, we find $y_{m,n} = 0$ for every $|m| > n$ and so the $n$th order
correction to $y$ comprises only $2n+1$ non-zero terms. As is common with such expansions, the complexity of the terms increases
rapidly with $n$. For instance, we find
$$
\lambda_2 = {\frac {\rho/g}{32}}\,{\frac { -3\,{\nu}^{6}+11\,{\nu}^{4}+15\,{\mu
}^{2}{\nu}^{4}-35\,{\mu}^{2}{\nu}^{2}+4\,{\nu}^{2}+8\,{\mu}^{2} }{  {\nu}^{3} \left( {\nu}^{2}-1 \right)^{3} \left( {\nu}^{2}-4 \right)}}
$$
whilst $\lambda_3$ is more complicated still.  By using {\tt Maple} or some other  symbolic manipulation software, one may nevertheless
compute quickly the first dozen terms. Our computations suggest that the series has a strictly positive--- although finite--- radius of convergence,
say $R=R(\mu,g,\rho)$. Given $v \ne 2g$, we can therefore compute a good approximation of the integrated density of states $N(E)$ in the range
$$
0 < E <  \frac{R g^2}{| 2 g/v -1|}
\,.
$$
This is illustrated in Figure~\ref{fig:Aurelien}.
Furthermore, this approximation becomes exact in the limit $E \rightarrow 0$. When we use the well-know asymptotics
of the Hankel function, we see that the upshot of the formulae (\ref{firstOrderCorrectionForY})-(\ref{firstOrderCorrectionForLambda}) is that the power law behaviour (\ref{exactLowEnergyBehaviour}) holds--- not only for $v=2g$--- but also for every $v>0$.

We remark however that the procedure only makes sense when $\nu$ is not a natural number. The origin of this restriction is obviously the identity (\ref{figueiredoRelation}).
We shall return to this point in \S \ref{sec:Conclusion}.

Let us now use the expansion to work out the low-energy behaviour of $\Omega$.
By using the formula,
$$
H_{\beta}^{(1)} (z) \simeq - \frac{\I}{\pi} \frac{\Gamma(\beta)}{(z/2)^\beta} \hspace{0.25cm} \mbox{as $z \rightarrow 0$}
$$
we readily deduce
\begin{eqnarray}
y(k/g) &\simeq& - \frac{\I}{\pi}\sum_{n=0}^\infty \varepsilon^n \sum_{|m| \leq n} y_{m,n} \Gamma ( \lambda + 2 m) 2^{\lambda+2 m} (k/g)^{-\lambda-2m} 
\nonumber
\\
\nonumber
&\simeq& - \frac{\I}{\pi} \left ( \frac{k}{2 g} \right )^{-\nu} \sum_{n=0}^\infty \varepsilon^n \,y_{n,n} \,\Gamma (\nu+2n) \,4^n \left ( \frac{E}{g^2} \right )^{-n} \,.
\end{eqnarray}
That is:
\begin{equation}
y(k/g) \simeq
- \frac{\I}{\pi} \left ( \frac{k}{2 g} \right )^{-\nu} \sum_{n=0}^\infty y_{n,n} \Gamma (\nu+2 n) \left [ 4 ( 1- 2 g/v) \right ]^n \hspace{0.25cm}
\mbox{as $k \rightarrow 0$}\,.
\label{intermediateLowEnergySolution}
\end{equation}
A careful examination of the recurrence relation (\ref{threeTermRecurrence}) for the $y_m$ reveals that
\begin{equation}
\notag
y_{n,n} = - \frac{c_{n,1}}{b_{n,0}} y_{n-1,n-1} 
= - \frac14 \frac{(\nu+2n-2)^2-\mu^2}{(\nu+2n-2)(\nu+2n-1)}\,
\frac{y_{n-1,n-1}}{\nu^2-(\nu+2n)^2}
\:.
\end{equation}
Hence, it is easily shown by induction that
$$
y_{n,n} = \frac{\Gamma (\nu+1) \,\Gamma (\nu)}{\Gamma \left ( \frac{\nu+\mu}{2} \right )\,\Gamma \left ( \frac{\nu-\mu}{2} \right )}
\frac{1}{4^n n!} \frac{\Gamma \left ( \frac{\nu+\mu}{2} +n \right )\,\Gamma \left ( \frac{\nu-\mu}{2} +n \right )}{\Gamma (\nu+n+1)\,
\Gamma ( \nu+2 n)}\,.
$$
When we report this in Equation (\ref{intermediateLowEnergySolution}), we recognise the familiar hypergeometric series:
\begin{equation}
y(k/g) \simeq - \frac{\I}{\pi} \left ( \frac{k}{2g} \right )^{-\nu}\,\Gamma (\nu) \,_2 F_1 \left ( \frac{\nu+\mu}{2}, \frac{\nu-\mu}{2};\,1+\nu;\, 1- 2g/v \right )\,.
\label{lowEnergySolution}
\end{equation}

We are now ready to examine the complex Lyapunov exponent~:
from Equation (\ref{complexLyapunovExponentAgain}), we then deduce that
\begin{equation}
\gamma (E) \underset{E\to0+}{\simeq} \nu g + 2 g (1- 2g/v) 
\frac{_2 F_1 ' \left ( \frac{\nu+\mu}{2}, \frac{\nu-\mu}{2};\,1+\nu;\, 1- 2g/v \right )}{_2 F_1 \left ( \frac{\nu+\mu}{2}, \frac{\nu-\mu}{2};\,1+\nu;\, 1- 2g/v \right )} 
\,.
\label{lowEnergyLyapunov}
\end{equation}
This agrees with our earlier finding (\ref{eq:Omega0SolvableModel})~; the form (\ref{lowEnergyLyapunov}) may also be obtained by injecting directly (\ref{eq:MixedZeroEnergyFofQ}) in (\ref{eq:RelationFPrimeOmega}).
Turning next to the IDOS, using (\ref{eq:IDoSfctY}), we deduce
\begin{equation}
\label{lowEnergyDensityOfStates}
  \boxed{ 
  N(E) 
  \underset{E\to0+}{\simeq} \frac{2g}{\left [ \Gamma (\nu) \,_2F_1 \left (\frac{\nu+\mu}{2},\,\frac{\nu-\mu}{2};\,\nu+1; 1-2g/v \right ) \right ]^{2}} 
  \left(\frac{E}{4g^2}\right)^\nu
  }
\end{equation}
We have thus shown that the occurence of the power law with exponent $\nu=\sqrt{\mu^2+2\rho/g}$ is a property
of the integrated density of states which holds over the full range of the values of the four parameters $\mu$, and $g,\,\iw,\,\rho$.
Interestingly, the prefactor in (\ref{lowEnergyDensityOfStates}) involves the same hypergeometric function as in the solution for $E=0$; see Eq.~(\ref{eq:Omega0SolvableModel}).
This is no coincidence. We shall return in due course to the relationship between the zero-energy
solution and the prefactor.

\subsection{Arbitrary distribution of the weights}
\label{arbitrarySubsection}

Let us now explain briefly how the results obtained for the particular case of exponentially-distributed weights
may be extended to more general distributions with support on $\mathbb{R}_+$. When we make the subsitution
(\ref{substitution}) in the equation (\ref{eq:EqDiffMixedCase}), we find, after some re-arrangement,
\begin{equation}
y''(x) + \frac{1}{x} y'(x) + \left ( 1 - \frac{\nu^2}{x^2} \right )\,y(x) = U \left ( \frac{g x^2}{2 k^2} \right ) \, y(x)
\label{generalEquationForY}
\end{equation}
where 
$$
U (\eta) := \int_0^{\infty} \e^{- t \eta} I'(t)\,\D t
\,,\hspace{0.5cm}
I (t) := \frac{2 \rho}{g} \e^{\frac{t}{2 g}}\,\int_t^\infty p( \tau)\,\text{d} \tau
$$
and $p$ is the probability density of the weights. It is clear that the right-hand side of this differential equation becomes
negligible in the low-energy limit. Indeed, if we assume for convenience that $p$ is smooth at zero, the coefficient of $y$ in
this right-hand side has
the asymptotic expansion
$$
U \left ( \frac{g x^2}{2 k^2} \right )  \simeq 
\sum_{\ell=1}^\infty \left ( \frac{2 k^2}{g} \right )^\ell I^{(\ell)} (0) \,x^{-2 (\ell+1)} \quad \text{as $k \rightarrow 0$}\,.
$$
So the ``perturbation'' on the right-hand side of Equation (\ref{generalEquationForY}) is, in the language of scattering theory used in Ref.~\cite{Cav94},
again ``long-range''. 
We may then, as in the exponentially-distributed case, seek a solution of the form (\ref{hankelSeries}).
Furthermore, since the perturbation is expressible in terms of inverse powers of $x^2$, we shall again be able
to make use of the contiguity relation (\ref{figueiredoRelation}) for Bessel functions~; by iteration, we find
$$
\frac{1}{x^{2 \ell}} H_{\beta}^{(1)} (x) = \sum_{|j| \leq\ell} c_{j,\ell} (\beta) \,H_{\beta+2j}^{(1)} (x)
$$
where the coefficients $c_{j,\ell}$ in this identity satisfy a first-order recurrence relation in the index $\ell$ and are therefore
easy to compute.
Substitution of the Hankel series (\ref{hankelSeries})  in (\ref{generalEquationForY}) then yields
\begin{multline}
\left [ (\lambda+2m)^2 - \nu^2 \right ] y_m 
= \sum_{\ell=1}^\infty \left ( \frac{2 k^2}{g} \right )^\ell I^{(\ell)} (0) \sum_{|j| \leq\ell}
c_{j,\ell} \left ( \lambda+2 m - 2 j \right ) \,y_{m-j}\,.
\notag
\end{multline}
By contrast with the exponentially-distributed case, the order of this recurrence relation for the $y_m$ is now {\it infinite}. But we may again
expand $\lambda$ and $y_m$ in powers of the small parameter $k^2$.
At each order in $k^2$, only a finite 
number of terms on the right-hand side need be retained, and the calculation of the coefficients
can proceed as in the exponentially-distributed case. In particular, 
we find for $y$ and $\lambda$ the same asymptotic
expressions as in Equations (\ref{firstOrderCorrectionForY}) and (\ref{firstOrderCorrectionForLambda}), except that
$\varepsilon$ is now replaced by
$$
\frac{k^2}{g^2} \left [ 1 - 2 g \,p(0) \right ]\,.
$$
Furthermore
$$
y(k/g) \simeq C k^{-\nu} \quad \text{as $k \rightarrow 0$}
$$
where the constant $C$ is an infinite sum involving the coefficients $y_{n,n}$, and thus, by Equation (\ref{eq:IDoSfctY}),
\begin{equation}
N(E) \simeq \frac{2 g}{\left ( \pi C \right )^2} E^\nu \quad \text{as $E \rightarrow 0+$}\,.
\label{lowEnergyIDOS}
\end{equation}


\section{Summary and concluding remarks}
\label{sec:Conclusion}
In this final section, we summarise our main findings and elaborate their implications for the long-time behaviour of a diffusion in a random
environment with absorption.

\subsection{Disordered quantum mechanics}

In this article, we have studied the spectral and localisation properties of a quantum Hamiltonian made up of two parts:
a ``supersymmetric'' part with superpotential $\phi = \Phi'$, where $\Phi$ is a Brownian motion of variance $g$ and drift $\mu g$,
 and a ``scalar'' potential $\Absor = K'$, where $K$ is a compound Poisson process of intensity $\rho$ whose weights (i.e. jumps) are positive, have
 probability density $p$ and mean $v$. Our analysis used the Dyson--Schmidt method and therefore focused for the most part
 on finding the decaying solution of a certain differential equation,  Equation (\ref{eq:FdeQ}), in terms of which one can express the complex Lyapunov
 exponent of the disordered system.
 
\begin{figure}[!ht]
\centering
\includegraphics[scale=0.9]{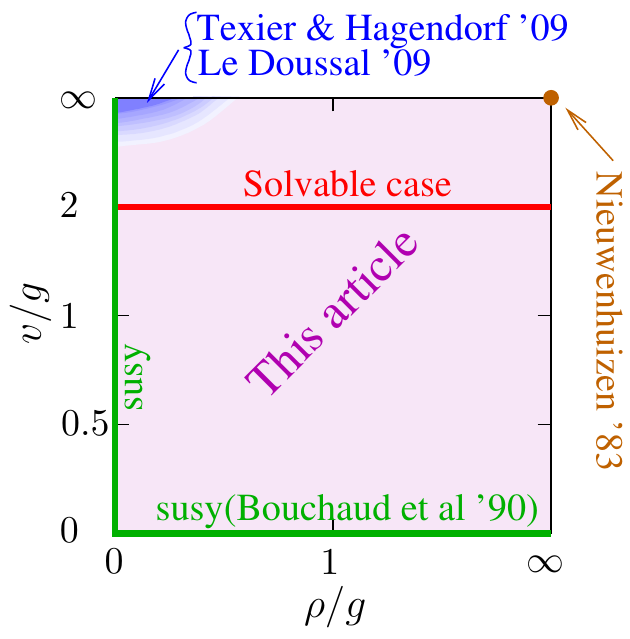}%
  \caption{The characteristic function $\Omega=\gamma-\I\pi N$ for the model (\ref{eq:hamiltonian}), when $\phi(x)$ is a Gaussian white noise of strength $g$ and  $\Absor(x)$ is a Poissonnian white noise with exponentially-distributed $\delta$ impurity weights for a mean value $\iw$, is known in several limits.
}
  \label{fig:PhaseDiagram}
\end{figure}

We have found a new solvable case~: that where $p$ is an exponential density with mean $v=2g$. It turns out that the complex Lyapunov
exponent is then the same as in the purely supersymmetric case (i.e. with $\Absor=0$), except for a shift in the drift of $\Phi$:
$$
  \mu\to\nu := \sqrt{\mu^2+2\rho/g}
  \:.
$$ 
We have given an alternative derivation of this striking fact  that uses the replica trick~\cite{Com10}.
One immediate consequence is the power law behaviour of the integrated density of states at low energy~: 
\begin{equation}
  \label{eq:powerLaw}
  \boxed{
  N(E) 
  \underset{E\to0+}{\simeq} \frac{2g}{\big[ \CsteYves \,\Gamma (\nu)  \big]^{2}} 
  \left(\frac{E}{4g^2}\right)^\nu  
  }
\end{equation}
where $\CsteYves=1$ for $v=2g$.
We have then shown, by solving the differential equation in terms of a Hankel series,
that this power law behaviour holds whatever the mean $v$ of the exponential distribution--- the constant $\CsteYves$ in the prefactor being 
now given by
$$
\CsteYves  =  {_2}F_1 \left (\frac{\nu+\mu}{2},\,\frac{\nu-\mu}{2};\,\nu+1; 1-2g/v \right )\,.
$$
Strictly speaking, our method of solution in this case is valid only for non-integral values of $\nu$. For the solvable case, there is of course
no such restriction, and our evidence is that the restriction is a technical artefact connected with the peculiarities of Bessel functions.
Indeed, we note that, in the solvable case,
the Lyapunov exponent exhibits some logarithmic dependence for integral values of the index (Section~\ref{supersymmetricDisorderSubsection}); this might be the cause of
the breakdown of the \og perturbative \fg{} approach of Section~\ref{sec:HankelSeries}.
Finally, we have extended the validity of the power law to a general probability density $p$ supported on ${\mathbb R}_+$. 
We have not been able to compute the prefactor explicitly, but we have strong reasons to believe that Equation (\ref{eq:powerLaw}) holds with $F_0(q)$ is the unique zero-energy ($E=0$) solution of our fundamental equation (\ref{eq:FdeQ}) with the asymptotic behaviour
$$
F_0(q) \simeq  \left ( 1+ 2 \I g q \right )^{\frac{\mu-\nu}{2}} \quad \text{as $q \rightarrow \infty$}
\,;
$$
Our analysis therefore provides a significant extension of the results obtained in earlier works where only limiting cases
could be treated successfully \cite{TexHag09,LeD09}.  The range of applicability of these different works is summarised diagramatically in Figure~\ref{fig:PhaseDiagram}.
Table~\ref{tab:SingularitesSpectrales} lists several known low-energy behaviours
of the spectral density $\varrho(E) = N'(E)$ of one-dimensional disordered Hamiltonians.

\begin{table}[!ht]
  \centering
  \begin{tabular}{lll}
    Model & Density of States  & Reference \\ \hline\hline\hline
    & & \\
    $H=-\deriv{^2}{x^2}+\Absor$  && \\
    & & \\ \hline\hline
    $\Absor$ Gaussian  
    & $\varrho(E\to-\infty)\sim|E|\exp\{-\frac8{3\sigma}|E|^{3/2}\}$ 
    &  \cite{Hal65}
    \\ \hline
    $\Absor$ Poissonian,  $\rho\to0$ 
    & $\varrho(E\to0^+)\sim E^{-3/2}\exp\{-\pi\rho E^{-1/2}\}$
    &  \cite{FriLlo60,Lif65}
    \\\hline
    $\Absor$ Poissonian, $\rho\to0$,  
    & $\varrho(E\to-v^2/4)\sim|E+v^2/4|^{-1+2\rho/v}$ 
    & \cite{Hal67} 
    \\
    $p(v_n)= \delta(v_n+v)$, $v>0$
    &
    & \\\hline
    $\Absor$ Poissonian, $\rho\to0$,
    & $\varrho(E\to-\infty)\sim|E|^{-1/2}\exp\{-(2/v)|E|^{1/2}\}$ 
    & \cite{ComTexTou10}
    \\
    $p(v_n)=(1/v)\e^{v_n/v}$, $v_n<0$ 
    &
    & 
    \\\hline\hline\hline
    & & \\
    $H=-\deriv{^2}{x^2}+\phi^2+\phi'$ &&  \\
     & &  \\\hline\hline
    $\phi$ Gaussian, $\mphi=0$
    & $\varrho(E\to0^+)\sim E^{-1} |\ln E|^{-3}$ 
    & \cite{Dys53,BouComGeoLeD90} 
    \\ \hline
    $\phi$ Gaussian,  $\mphi=\mu\,g$
    & $\varrho(E\to0^+)\sim E^{-1+\mu}$ 
    & \cite{BouComGeoLeD90} 
    \\ \hline
    $\phi$ Poissonian
    & $\varrho(E\to0^+)\sim E^{-\frac{3v+1}{2v}}\exp\{-\frac\pi2\rho E^{-1/2}\}$ 
    & \cite{Bie08,ComTexTou10}  
    \\
    $p(v_n)=(1/v)\e^{-v_n/v}$, $v_n>0$ 
    &
    & 
    \\\hline\hline\hline
     & & \\
    $H=-\deriv{^2}{x^2}+\phi^2+\phi'+\Absor$ && \\
    & & \\\hline\hline
    $\Absor$ and  $\phi$ Gaussian of & & \\
    variance $\sigma$ and $g$, $\mphi=0$
    & $\varrho(E\to-\infty) \sim \exp\{-\frac{\pi}{\sqrt{\sigma g}}|E|\}$ 
    &  \cite{HagTex08}
    \\\hline
    $\phi$ Gaussian, $\Absor$ Poissonian & & \\
    $v \rightarrow \infty$, $\rho \to 0$, $\mphi =0$
    & $\varrho(E\to0^+)\sim E^{-1+\sqrt{2\rho/g}}$
    &  \cite{TexHag09} \\
     $v \rightarrow \infty$, $\rho \to 0$, $\mphi =\mu g$
    & $\varrho(E\to0^+)\sim E^{-1+\sqrt{\mu^2+2\rho/g}}$
    & \cite{LeD09}  \\
    $v_n>0$ 
    & $\varrho(E\to0^+)\sim E^{-1+\sqrt{\mu^2+2\rho/g}}$
    & This article
    \\\hline\hline
    \\
  \end{tabular}
  \caption{\it Low-energy behaviour of the density of states
    $\varrho(E)=N'(E)$ for various disordered models 
    where $\Absor$ and $\phi$ are (Gaussian and/or Poissonian) white noises. Here, ``Poisonnian'' means that the noise arises from a compound Poisson process
   of intensity $\rho$ with a probability density $p$ of the weights $v_n$ with mean $v$.}
  \label{tab:SingularitesSpectrales}
\end{table}

For $\mu=0$, in the limit 
$\rho \rightarrow 0$ and $v \rightarrow \infty$, we can, following  \cite{TexHag09}, explain the occurrence
of the exponent $\nu=\sqrt{2\rho/g}$ by adapting the heuristic Lifshits argument that we outlined
in \S \ref{sec:LimitBehaviourScalar} when we discussed the monolithic scalar case.
The occurrence of low-energy states is related to the occurrence of large intervals of width $\ell$ free of impurities via the asymptotic formula
\begin{equation}
  \label{eq:14}
  \varrho(E) \sim \int_0^\infty\D\ell\,\rho\,\e^{-\rho\ell} 
  W_0(E;\ell) \quad \text{as $E \rightarrow 0^+$}\,.
\end{equation}
In this expression, $W_0(E;\ell)$ is the distribution of the ground state energy of the supersymmetric Hamiltonian (\ref{eq:Hsusy}) on a finite interval of length $\ell$
with Dirichlet boundary conditions.
We then use the known fact that $W_0$ has a log-normal tail~\cite{Tex00,TexHag10}, i.e.
$$
 W_0(E;\ell)\sim\frac1E\exp-\frac{1}{2g\ell}\ln^2(g^2/E) \quad \text{as $E \rightarrow 0^+$}\,.
$$ 
The integral on the right-hand side of Equation (\ref{eq:14}) can be estimated by the steepest descent method~; the dominant contribution
comes from the vicinity of  $\ell_*=\frac1{\sqrt{2\rho g}}\ln(g^2/E)$, and this leads to the power law~(\ref{eq:powerLaw}) with $\mu=0$.

Mathematically, the robustness of the power law may be understood as follows. The effect of the scalar noise $\kappa$
on the low energy behaviour of the solution $\hat{f}$ of our fundamental
equation (\ref{eq:FdeQ}) is, in the first approximation, completely determined by the large $q$ behaviour of the L\'{e}vy exponent. For a compound Poisson
process of intensity $\rho$, the
L\'{e}vy measure $m$ is finite, and $m({\mathbb R}) = \rho$. Hence
$$
\frac{\Lambda ( q)}{q} \simeq \frac{\rho}{q} \quad \text{as $q \rightarrow \infty$} \,. 
$$
The point is that the right-hand side does not depend on 
the probability density $p$ of the weights. The situation would be very different if the absorption process were a subordinator with an {\it infinite} L\'{e}vy measure, as for the examples discussed in Appendix \ref{appendix:SPLN}.
The analysis of the model \eqref{eq:hamiltonian} in this case would be of interest.

Another facet of our work is the analysis of the low-energy localisation properties of the disordered Hamiltonian (\ref{eq:hamiltonian}). An interesting feature occurs in the limit of low density $\rho\to0$ and large impurity weight $\iw\to\infty$. In this case, we found
 $$
 \gamma(0) \simeq \sqrt{2\rho g} + \rho\,\ln(v/2g)\,.
 $$
Starting from the supersymmetric (de-localized) case, the introduction of a weak concentration of impurities localizes the low energy states on a scale smaller than the typical distance between impurities $1/\gamma(0)\ll1/\rho$. Naively, one might have guessed the opposite or, say, $1/\gamma(0)\sim1/\rho$.
It would therefore be of interest to study the spatial structure of the low-energy eigenstates in this limit.

All the results we have reported here were obtained for a L\'evy process $\IntAbsor$ whose L\'evy measure $m(d y)$ has \textit{positive} support, so that the weights $v_n$ are necessarily positive.
If, instead, we allowed the noise $\Absor$ to take also negative values, then a fraction of the spectrum would spill unto $\mathbb{R}_-$. For example, the spectral density of the mixed model in the case where $\Absor$ is a Gaussian white noise was shown by Texier \& Hagendorf \cite{HagTex08} to exhibit an exponential behaviour at low energy (Table~\ref{tab:SingularitesSpectrales}). The nature of the spectral density of the model \eqref{eq:hamiltonian} when $\IntAbsor$ is a more general L\'evy process remains an open question.

\subsection{Sinai diffusion with absorbers}

Let us now revisit the problem that motivated this study--- that of determining the long-time behaviour of the return probability (\ref{returnProbability})
for a Sinai diffusion with random absorbers--- and discuss the implications
of our findings. 

Recall that the return probability in a driftless environment without absorption is given by
$\ReturnProba(t)=1/\sqrt{4\pi t}$ (free case). 
When considered separately, $\phi$ and $\kappa$ have a very different impact on the return probability:
the white noise $\phi$ (without absorption) slows down 
the diffusion,  leading to the extremely slow decay $\ReturnProba(t) \sim g/\ln^2(g^2t)$ for long times (Sinai). On the other hand, the introduction
of random absorbers in a driftless deterministic environment leads to the stretched exponential decay $\ReturnProba(t) \sim\exp\{-3(\frac\pi2)^{2/3}(\rho^2t)^{1/3}\}$ (Lifshits). 
The question of interest is therefore the following~: {\it When the random force field and the absorbers are combined, will the tail of the return probability
be heavier or lighter than in the free case~?}
By the usual Tauberian argument, Formula \eqref{returnProbability} and the low energy asymptotics $\rho(E) = N'(E) \sim E^{\nu-1}$ imply
\begin{equation}
  \label{eq:PowerLawP}
  \ReturnProba(t) \sim t^{-\sqrt{\mu^2+2\rho/g}}
  \hspace{1cm}
  \mbox{for } t\to\infty
  \:.
\end{equation}
Our results therefore show that the answer to this question
depends only the relative sizes of the density of absorbers $\rho$ and of the variance $g$ of the Brownian environment~: {\it the decay of the return probability is faster than for the free diffusion if $\rho>g/8$, and slower otherwise.}
In particular, the answer does not depend on the efficiency of the absorbing sites, which is encoded in the distribution of the weights $v_n$.
Le~Doussal has pointed out that the algebraic decay \eqref{eq:PowerLawP} is largely the result of absorption since the survival probability has---
save for a  logarithmic correction--- the same power law \cite{LeD09}.

We summarize in table~\ref{tab:ReturnProb} the range of behaviours exhibited by the return probability for various diffusions.

\begin{table}[!ht]
  \centering
  \begin{tabular}{llll}
    Diffusion & Long time behaviour   &\\ \hline\hline\hline
    With absorbtion & &\\ \hline\hline
    $\Absor$ Poissonian,  $\rho\to0$ 
    & $\ReturnProba(t) \sim t^{1/6}\exp\{-3(\frac\pi2)^{2/3}(\rho^2t)^{1/3}\}$
    &(Lifshits) 
    \\\hline\hline\hline
    In a random force field &&&    \\\hline\hline
    $\phi$ Gaussian, $\mphi=0$
    & $\ReturnProba(t)\sim g/\ln^2(g^2t)$ 
    &(Sinai)
    \\ \hline
    $\phi$ Gaussian, $\mphi=\mu\,g$
    & $\ReturnProba(t) \sim t^{-\mu}$  
    &\\ \hline
    $\phi$: Poissonian,
    &  
    &\\
    $p(v_n)=1/v\,\e^{-v_n/v}$, $v_n>0$
    & $\ReturnProba(t) \sim t^{\frac16+\frac{1}{3v}} \,\exp\{-3(\frac\pi4)^{2/3}(\rho^2t)^{1/3}\}$ 
    &\\\hline\hline\hline
    Mixed case && \\\hline\hline
    $\phi$ Gaussian,  $\mphi=\mu\,g$,
    & $\ReturnProba(t) \sim t^{-\sqrt{\mu^2+2\rho/g}}$ 
    & 
    \\ 
    $\Absor$ Poissonian,  $v_n>0$
    &
    & 
    \\\hline\hline
    \\
  \end{tabular}
  \caption{\it Long-time behaviour of the return probability for diffusions
  with a concentration $\rho$ of absorbers of effectiveness $v_n$ and/or a random force field of strength $g$ and drift $\mu g$. The results are deduced
  from Table~\ref{tab:SingularitesSpectrales}. }
  \label{tab:ReturnProb}
\end{table}

Although we have succeeded in extracting from the Dyson-Schmidt method results for the density of states and the Lyapunov exponent of the model (\ref{eq:hamiltonian}) which are largely free of restrictions on the parameters, it is fair to say that these two quantities provide only rather basic information on the behaviour of
the diffusion characterised by the generator \eqref{generator}. By contrast, 
by using a real space renormalisation group method, Le~Doussal~\cite{LeD09} was able to obtain approximations--- valid in certain asymptotic limits--- of many other 
interesting quantities and objects such as the averaged survival probability, the averaged propagator of the diffusion, the distribution of the traveled distance before absorption. The obvious but difficult question is whether the Dyson--Schmidt approach may be adapted so as to recover or extend
more of the results found by Le Doussal.

\section*{Acknowledgements}

AG thanks the University of Bristol for its hospitality while part of this work was carried out. 
CT would like to thank Christian Hagendorf, with whom his interest in this topic began,  as well as Alberto Rosso for some helpful suggestions. 
YT is grateful to the Laboratoire de Physique Th\'{e}orique et Mod\`{e}les Statistiques for generously funding his visits to Orsay during the project. 
We also thank Alain Comtet for many stimulating discussions, and for allowing us to include in this paper the calculation presented in~\S~\ref{subsec:Alain}.


\begin{appendix}

\section{The generating functional of a L\'evy noise}
\label{appendix:SPLN}

The purpose of this appendix is to discuss some statistical properties of L\'evy processes
and to make the connection between two useful concepts~: the L\'evy exponent $\Lambda$
and the generating functional of the corresponding noise, used in section~\ref{subsec:Alain}.

\subsection{Generating functional and L\'evy exponent}

Given a L\'{e}vy process $L(x)$, a convenient way to encode the statistical properties of the noise $L'(x)$ is through the generating functional $W[h]$, defined as 
\begin{equation}
  \e^{W[h]} := \mean{\e^{\int\D x\,h(x)\,L'(x)}}
  \:.
\end{equation}
By partitioning the domain of integration into subintervals of length $\epsilon>0$, we can write
\begin{eqnarray}
  \e^{W[h]} &=& \lim_{\epsilon\to0}\mean{\prod_{n} \e^{\int_{n\epsilon}^{(n+1)\epsilon}\D x\,h(x)\,L'(x)}}
  =\lim_{\epsilon\to0} \prod_{n}\mean{\e^{h(n\epsilon)\,[L((n+1)\epsilon)-L(n\epsilon)]}}
  \nonumber\\
  &=& \lim_{\epsilon\to0}\prod_{n} \e^{-\epsilon\,\Lambda(\I h(n\epsilon))}
  =\e^{- \int\D x\, \Lambda( \I h(x)) }
  \:,
\end{eqnarray}
where the second equality has used the Markov property of the L\'evy process.
This shows that 
\begin{equation}
  \boxed{ 
   W[h] =  - \int\D x\, \Lambda( \I h(x)) 
   }
\end{equation}

\begin{example} 
Consider the compound Poisson process
$$
L(x) = \sum_{j=1}^{\poisson(x)} Y_j
$$
where $\poisson(x)$ is a Poisson process of intensity $\rho$ and the $Y_j$ are independent and distributed with the same exponential density
$p(y)=(1/v)\,\e^{-y/v}$ for $y>0$, i.e.
$\hat p(\I q)=\frac1{1-qv}$.
The jumps $Y_j$ have moments $\mean{Y^n}=n! v^n$ and the generating functional is therefore given by
\begin{equation}
  \label{eq:GFCPP}
   W[h]  = \rho \int\D x\, \frac{h(x)}{1/v- h(x)}
  \:.
\end{equation}
\end{example}

\subsection{Correlation functions}

Knowledge of the generating functional simplifies the analysis of the connected correlation function of the L\'{e}vy noise $L'(x)$:
\begin{eqnarray}
\label{eq:CumulantsLevyNoise}
\cum{L'(x_1)\cdots L'(x_n)}
  &=& \derivf{^nW[h]}{h(x_1)\cdots\delta h(x_n)}\Big|_{h=0} 
  \\
  &=& -\I^n\Lambda^{(n)}(0)\,\,\delta(x_1-x_2)\cdots\delta(x_1-x_n)
\end{eqnarray}
where we assumed that the L\'evy exponent is $n$ times differentiable. (This is equivalent to assuming the
existence of the first $n$ moments of the L\'evy process.)
The derivatives of the L\'evy exponent may be expressed in terms of the cumulants of the process:
\begin{equation}
  -\I^n\Lambda^{(n)}(0) = \cum{L(1)^n} \,.
\end{equation}
For a compound Poisson process,  they may be expressed also in terms to the \textit{moments} of the $Y_j$:
$$
\cum{L(1)^n}=\int m(d y)\,y^n\,.
$$
By integrating (\ref{eq:CumulantsLevyNoise}) over the $n$ coordinates, we deduce the connected correlation function of the L\'{e}vy process $L(x)$ itself~: 
  \begin{equation}
    \label{eq:CorrelationsLevyProcess}
    \cum{L(x_1)\cdots L(x_n)}
    = \cum{L(\mathrm{min}\{x_1,\cdots,x_2\})^n}
    =\mathrm{min}\{x_1,\cdots,x_2\}\, \cum{L(1)^n}
    \:.
  \end{equation}
  This reflects the Markovian nature of the process.

\subsection{Infinite L\'evy measures}

There are L\'{e}vy  subordinators whose L\'{e}vy measure $m$ is infinite, i.e. $m({\mathbb R}_+) = \infty$. By Equation (\ref{levyCondition}) this can only
happen if the measure is sufficiently singular at the origin. Such processes are not in the class of compound Poisson processes.
We proceed to discuss  two examples which will make it clear that the singular nature of the measure affects the large $q$ behaviour
of the L\'{e}vy exponent.

\begin{example}
The Gamma subordinator
has L\'{e}vy measure $m(d y)= (\D y/y)\,\e^{-y/v} $ for $y>0$ and L\'evy exponent $\Lambda(q)=\ln(1+\I v q)$. The singular nature of the measure is reflected in the logarithmic  growth of the exponent as $q\to\infty$. The cumulants are given by  $\cum{L(1)^n}=(n-1)! v^n$.
\end{example}

\begin{example}
The alpha-stable subordinator has L\'{e}vy measure $m(d y)=\frac{\alpha}{\Gamma(1-\alpha)}\frac{\D y}{y^{1+\alpha}}$ for $y>0$ and  $0<\alpha<1$.
The corresponding L\'evy exponent is $\Lambda(q)=(\I q)^\alpha$. The L\'evy exponent has a branch point at $q = 0$ and, because of 
the power law tail of the L\'evy measure,
the moments 
$\overline{L(x)^n}=\infty$ are infinite.
The simple power-law form of the L\'evy  exponent however leads to the scaling $L(x)\eqlaw x^{1/\alpha} L(1)$~; see \cite{BouGeo90,ComTexTou11}.
\end{example}


\section{The Green's function, the functional determinant and a Thouless formula}
\label{app:Thouless}

The distribution of the Riccati variable has some interesting connections with the 
 averaged Green's function
$$
G(E) := \mean{\bra{x}\frac{1}{E-\mathscr{H}}\ket{x}}\,.
$$
That $G$ is independent of $x$ follows from the fact that the disorder is translation-invariant.
We have seen in \S  \ref{sec:FdeQ} that the left and right derivatives of the Fourier transform $\hat{f}(q)$ may be expressed in terms
of the complex Lyapunov exponent just below or above the spectrum:
$$
\hat{f}'(0^\pm)=\I[\mphi-\Omega(E\pm\I0^+)]\,.
$$
Using the analyticity of $\Omega(E)$ and the well-known relation between the density of states per unit length and the trace of the Green's function, $\pi N'(E)=\mp \im G(E\pm\I0^+)$, we may write
\begin{equation}
  \I \deriv{}{E} \hat{f}'(0^\pm) = \deriv{}{E} \Omega(E\pm\I0^+) =  G(E\pm\I0^+)
  \:.
\end{equation}
We may then express the complex Lyapunov exponent as a functional determinant. Indeed, writing 
$$
G(E) = \lim_{\text{Vol}\to\infty}\frac{1}{\text{Vol}}\mathrm{tr}\left\{\frac1{E-\mathscr{H}}\right\}
$$
where $\text{Vol}$ is the volume of the system, we obtain the formal representation
\begin{equation}
   \label{eq:SpectralDeterminant}
   \Omega(E)=\lim_{\text{Vol}\to\infty}\frac{1}{\text{Vol}} \ln\det(E-\mathscr{H}) + c
   \:,
\end{equation}
where $c$ is a constant independent on $E$. 
Functional determinants have attracted a lot of attention~; see for example \cite{GelYag60,For87}, or \cite{KirLoy08} for a pedagogical presentation. They have also stimulated several works on metric graphs~; see review articles in \cite{AkkComDesMonTex00,ComDesTex05}.
In practice, the calculation of a spectral determinant requires some regularization scheme. For example, using $\zeta$-regularization, it is known that the determinant of the operator acting on functions defined on $[0,\ell]$, satisfying Dirichlet boundary conditions is 
\begin{equation}
  \det(E-\mathscr{H})=2\,\psi(\ell;E) 
\end{equation}
where $\psi$ is solution of the Cauchy problem~(\ref{cauchyProblem})~; 
see for example \cite{HarKirTex12}.

Using the representation (\ref{eq:SpectralDeterminant}), we can among other things derive the Herbert-Jones-Thouless formula~\cite{HerJon71,Tho72,Luc92} as follows. Start from
\begin{equation}
  \frac{\det(E-\mathscr{H})}{\det(E'-\mathscr{H})} = \frac{\psi(\ell;E)}{\psi(\ell;E')}
  \:.
\end{equation}
Take the logarithm on both sides and consider the limit $\ell \to\infty$. The resulting left-hand side may be expressed in terms of the (bulk) density of states per unit length:
$$
\lim_{\ell\to\infty}\frac{1}{\ell}\ln\det(E-\mathscr{H})=\int\D\omega\,N'(\omega)\ln(\omega-E)\,.
$$ 
The resulting right-hand side is, by definition, the Lyapunov exponent (\ref{eq:DefLyapunov}). 
Integration by parts then yields 
\begin{equation}
  \label{Eq:Thouless1972}
  \gamma(E) - \gamma(E') = -\dashint\D\omega\,N(\omega)\,
  \left( \frac{1}{\omega-E} - \frac{1}{\omega-E'} \right)
  \:.
\end{equation}
In particular, setting $E'=0$ leads to
\begin{equation}
  \label{Eq:Thouless1972bis}
  \gamma(E) = \gamma(0)  - E\, \dashint\D\omega\,\frac{N(\omega)}{\omega(\omega-E)}
  \:.
\end{equation}

As an illustration, let us apply this formula to the supersymmetric Hamiltonian considered at the end of section~\ref{sec:FdeQ}, whose IDOS
exhibits the low energy behaviour $N(E)\sim E^\mu$. 
Rewrite (\ref{Eq:Thouless1972bis}) in the form 
$$
\frac{\gamma(E)-\gamma(0)}{E}=-\dashint_0^\infty\D\omega\,\frac{N(\omega)}{\omega(\omega-E)}\,.
$$
For $\mu<1$, the integral is dominated by the small $\omega$ as $E\to0^-$. Therefore 
$$
\gamma'(E)\sim\int_{|E|}^\infty\D\omega\,\frac{N(\omega)}{\omega^2}\sim(-E)^{\mu-1}\,.
$$
We conclude that $\gamma(E)\simeq\gamma(0)+a_\mu(-E)^\mu$ for $E\to0^-$, where $a_\mu$ is some constant. 

Alternatively, the relation between the Lyapunov exponent and the IDOS 
may be deduced from the Kramers-Kronig (Plemelj) formula, which exploits the analyticity of the complex Lyapunov exponent \cite{Luc92}.


\section{The ratio of hypergeometric functions (\ref{eq:Omega0SolvableModel})}

The behaviour of the ratio
$$
\frac{ 
              _2F_1\left(\frac{\nu}{2} , 1+ \frac{\nu}{2} ; 1+\nu ; 1 - 2g/\iw \right)
            }{
            _2F_1\left(\frac{\nu}{2} , \frac{\nu}{2} ; 1+\nu ; 1 - 2g/\iw \right)
            } 
$$
in the limit $\iw\to\infty$ can be approximated by using the integral representation of the hypergeometric function~\cite{gragra}. 
We see that the denominator reaches a finite limit. Setting $\epsilon=2g/\iw$~:
\begin{eqnarray}
  \nonumber
  B(\nu/2 , 1+\nu/2)\,
  &&_2F_1\left(\nu/2 , \nu/2 ; 1+\nu ; 1 - \epsilon \right)
  \\
  \nonumber
  &&=\int_0^1\D t\,t^{\nu/2-1}(1-t)^{\nu/2}\left[1-t(1-\epsilon)\right]^{-\nu/2}
  \underset{\epsilon\to0}{\longrightarrow}
  \frac{2}{\nu}
  \:.
\end{eqnarray}
The numerator diverges in the limit $\epsilon\to0$ since
\begin{eqnarray}
  \nonumber
  &&B\left(\frac\nu2 , 1+\frac\nu2\right)\,
  _2F_1\left(\frac\nu2 , 1 +\frac\nu2 ; 1+\nu ; 1 - 2g/\iw \right)
  \\
  \nonumber
  &&=\int_0^1\D t\,t^{\frac\nu2}
  (1-t)^{\frac\nu2-1}
  \left[1-t(1-\epsilon)\right]^{-\frac\nu2}
  =\int_0^1\D t\,(1-t)^{\frac\nu2}t^{\frac\nu2-1}\left[t+\epsilon-t\epsilon\right]^{-\frac\nu2}
  \:.
\end{eqnarray}
This last expression shows that, when $\epsilon\to0$, it is the neighbourhood of $t=0$ that makes the dominant contribution
to the integral. We deduce the estimate 
$$
    \epsilon^{-\nu/2}\int_0^\epsilon\D t\,t^{-1} 
  + \int_\epsilon^1\D t\,t^{-1}
  =\frac{2}{\nu} + \ln(1/\epsilon)
  \:.
$$
Hence
\begin{equation}
  \label{eq:RatioHFresult}
  \frac{ _2F_1\left(\nu/2 , 1+\nu/2 ; 1+\nu ; 1 - \epsilon \right) }
       { _2F_1\left(\nu/2 , \nu/2 ; 1+\nu ; 1 - \epsilon \right) } 
  \simeq 1 + \frac{\nu}{2}\ln(1/\epsilon) 
  \hspace{0.5cm}\mbox{as } \epsilon\to0 \mbox{ and } \nu\to0
  \:.
\end{equation}


\section{Mixed Gaussian disorder}

We consider in this appendix the case where $\phi$ and $\kappa$ are two independent Gaussian white noises, with the aim of re-deriving the result obtained by Texier \& Hagendorf \cite{HagTex08}
using the replica trick.
In the mixed Gaussian case, the differential equation (\ref{eq:FdeQ}) 
must be replaced by
\begin{equation}
  \left[
    -(1+2\I g\,q)\deriv{^2}{q^2} + 2\I g (\mu-1)\deriv{}{q}
    +E+ \frac{\I \sigma}{2}q
  \right] \hat{f}(q) = 0  
  \hspace{0.5cm}\mbox{ for }
  q>0
\end{equation}
where $\sigma$ is the variance of the Brownian process $K$.
This is a degenerate case of the hypergeometric equation. The change of variables~\cite{NikOuv83}
$$
  \hat{f}(q)  = z^\mu \e^{-z/2}\,y(z)\,,
  \quad
  z = \sqrt{\frac{\sigma}{g}}\,q-\frac\I2 \sqrt{\frac{\sigma}{g^3}} 
$$
transforms it into the Kummer equation
$$
  \left[
    z\deriv{^2}{z^2} + (\mu+1-z) \deriv{}{z}
    -a
  \right] y(z) = 0 
  \hspace{0.25cm}\mbox{where}\hspace{0.25cm}
  a := \frac{\mu+1}{2}
    -\I \, \varepsilon \hspace{0.25cm}\mbox{and}\hspace{0.25cm}
    \varepsilon :=  
         \frac{E}{2\sqrt{\sigma g}} - \frac18\sqrt{\frac{\sigma}{g^3}}
       \:.
$$
The solution decreasing at infinity may be expressed in terms of the Kummer function \cite{gragra}
\begin{equation}
  \hat{f}(q)  = c\,  z^\mu \e^{-z/2}\, \Psi\left(a,\mu+1;z\right)
 \:,
\end{equation}
where $c$ is a normalisation constant ensuring that $\hat{f}(0)=1$.

\begin{figure}[!ht]
  \centering
  \includegraphics[width=0.475\textwidth]{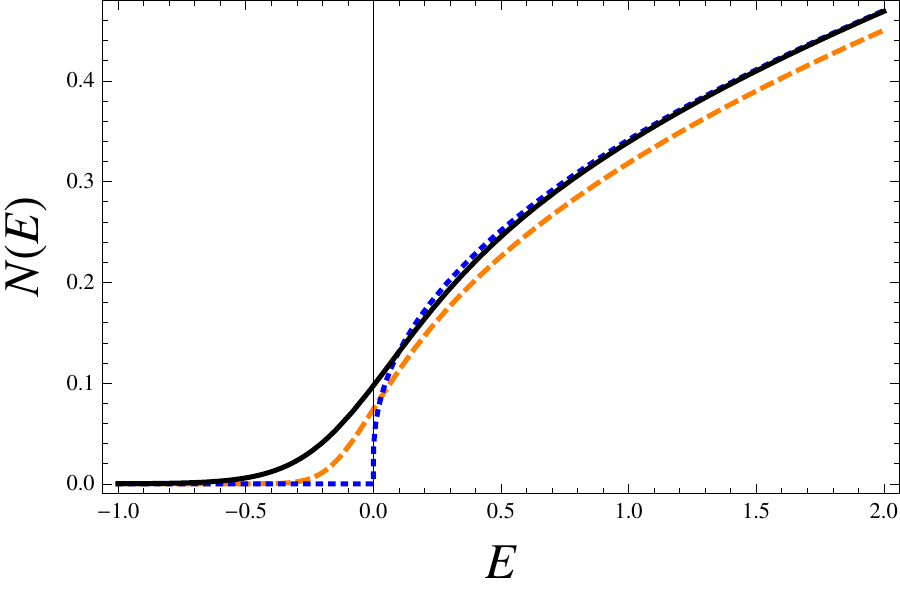}
  \hspace{0.25cm}
  \includegraphics[width=0.475\textwidth]{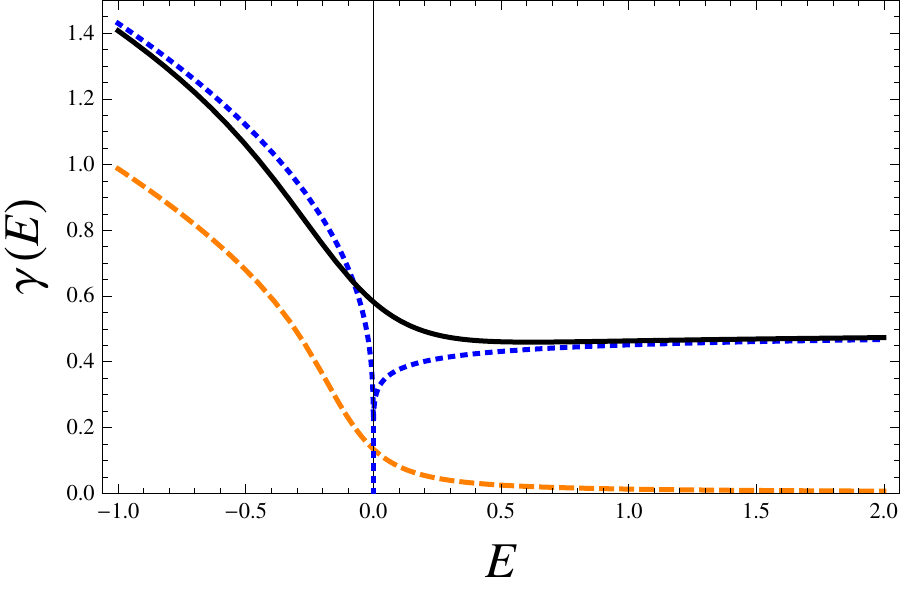}
  \caption{\it The IDOS and the Lyapunov exponent for the monolithic scalar Gaussian
    case (orange dashed), the monolithic supersymmetric Gaussian case (blue dotted) and the mixed Gaussian case (black continuous). 
    Here $g=1$, $\sigma=0.1$ and $\mu=0$. 
    }
  \label{fig:SusyGaussian}
\end{figure}

By using the identity $\Psi'(a,b;z)=-a\,\Psi(a+1,b+1;z)$, we obtain
\begin{equation}
  \label{eq:HagendorfTexier08}
  \Omega(E+\I0^+) = - \mu g 
  +g\,\zeta
  \left[
    1 + 2a\,
   \frac{\Psi\left(a+1,\mu+2;\zeta\right)}
        {\Psi\left(a,\mu+1;\zeta\right)}
  \right]
  \hspace{0.25cm}\mbox{for } \zeta := -\frac\I2 \sqrt{{\sigma}/{g^3}}
   \:.
\end{equation}
It seems fair to say that this derivation is more straightforward than that based on the replica trick and described 
in \cite{HagTex08}. We note in passing that Equation (102) of that
reference contains a misprint~: an additional $\mu+1$ at the
denominator. 
The result may also be expressed in terms of a Whittaker function~\cite{ComLucTexTou13}~: 
$$
  \Omega(E+\I0^+) = g \,
  \left( 
     1 - 2\zeta \,\frac{W'_{\I\varepsilon,\frac{\mu}{2}}(\zeta)}{W_{\I\varepsilon,\frac{\mu}{2}}(\zeta)}
  \right)
   \:.
$$

Hagendorf \& Texier also treated the case where $\Absor$ and $\phi$ are  {\it correlated} white noises;
see \cite{HagTex08}, Appendix A, where this case is mapped onto the uncorrelated case through a redefinition of the parameters. A more detailed discussion of the correlated case may also be found in Ref.~\cite{ComLucTexTou13}.

\end{appendix}



\end{document}